\documentclass[aps,prd,showpacs,twocolumn]{revtex4}
\usepackage{graphicx}
\usepackage{dcolumn}
\usepackage{bm}

\begin{document}

\def\be{\begin{equation}}
\def\ee{\end{equation}}
\def\bd{\begin{displaymath}}
\def\ed{\end{displaymath}}
\def\ba{\begin{eqnarray}}
\def\ea{\end{eqnarray}}
\def\lr{\leftrightarrow}
\def\nn{$n\bar n$ }
\def\qq{$q\bar q$ }
\def\cc{$c\bar c$ }
\def\ccbar{$c\bar c$ }
\def\x{\bf x}
\def\B{\rm B}
\def\D{\rm D}
\def\E{\rm E}
\def\F{\rm F}
\def\G{\rm G}
\def\H{\rm H}
\def\I{\rm I}
\def\J{\rm J}
\def\K{\rm K}
\def\L{\rm L}
\def\M{\rm M}
\def\P{\rm P}
\def\S{\rm S}
\def\T{\rm T}
\def\V{\rm V}

\title{\Large Higher Charmonia} 
\author{
T.Barnes,$^a$\footnote{Email: tbarnes@utk.edu}
S.Godfrey$^b$\footnote{Email: godfrey@physics.carleton.ca}
and
E.S.Swanson$^c$\footnote{On leave from the Department of Physics and Astronomy, 
University of Pittsburgh, Pittsburgh PA 15260. Email: swansone@pitt.edu}
}
\affiliation{
$^a$Department of Physics and Astronomy, University of Tennessee,
Knoxville, TN 37996,
USA,\\
Physics Division, Oak Ridge National Laboratory,
Oak Ridge, TN 37831, USA\\
$^b$Ottawa-Carleton Institute for Physics, 
Department of Physics, Carleton University, 
Ottawa K1S 5B6, Canada\\
$^c$Rudolph Peierls Centre for Theoretical Physics,\\
Oxford University, Oxford, OX1 3NP, UK.}

\date{\today}

\begin{abstract}
This paper gives results for the spectrum,
all allowed E1 
radiative partial widths 
(and some important M1 widths) 
and all open-charm strong decay amplitudes 
of all 40 $c\bar c$ states expected up to the mass of 
the 4S multiplet, just above 4.4~GeV.
The spectrum and radiative widths are evaluated using two models, 
the relativized Godfrey-Isgur model and a nonrelativistic potential model. 
The electromagnetic transitions 
are evaluated 
using Coulomb plus linear plus smeared hyperfine wavefunctions, 
both in a nonrelativistic
potential model and in the Godfrey-Isgur model. 
The open-flavor strong decay amplitudes are determined assuming
harmonic oscillator wavefunctions 
and the $^3$P$_0$ decay model. 
This work is intended to motivate future experimental studies of higher-mass 
charmonia, and may be useful for the analysis of 
high-statistics data sets to be accumulated by the
BES, CLEO and GSI facilities.

\end{abstract}

\maketitle

\section{Introduction}

Since its discovery in 1974 \cite{Aubert:1974js,Augustin:1974xw},
the charmonium system has become the prototypical 
`hydrogen atom' of meson spectroscopy 
\cite{Appelquist:1974zd,DeRujula:1974nx,Appelquist:1974yr,Eichten:1974af}.
The experimentally 
clear spectrum of relatively narrow states below the open-charm DD 
threshold of 3.73~GeV can be identified with 
the 1S, 1P and 2S \cc levels predicted by potential models, which incorporate 
a color Coulomb term at short distances
and a linear scalar confining term at large distances. Spin-dependent interquark 
forces are evident in the splittings of states within these multiplets, and
the observed splittings are consistent with the predictions of a 
one gluon exchange (OGE) Breit-Fermi Hamiltonian, combined with 
a linear scalar confining interaction. 
Discussions of the theoretical importance and experimental status of
heavy quarkonium, including recent experimental results for charmonium, 
have been given by
Quigg \cite{Quigg:2004nv},
Galik \cite{Galik:2004mi}, the CERN quarkonium working group
\cite{Brambilla:2004wf}, Seth \cite{Seth:2005pr,Seth:2005ai,Seth:2005ak}
and Swarnicki \cite{Skwarnicki:2005pq}.

Recently there has been a resurgence of interest in charmonium, 
due to the realization that B factories can contribute 
to the study of the missing \cc states 
\cite{Eichten:2002qv}, and to high-statistics experiments at
BES \cite{BES_report} and CLEO \cite{CLEO_report} 
and the planned GSI $p\bar p$ facility \cite{GSI_report}.

The possibility of contributions from B factories 
was dramatically illustrated by the recent discovery 
of the long missing $2^1{\rm S}_0$ $\eta_c'$ 
state by the Belle Collaboration \cite{Choi:2002na},
which has since been confirmed by BABAR \cite{Aubert:2003pt},
and has also been observed by CLEO in $\gamma\gamma$ collisions
\cite{Asner:2003wv}.

Additional interest in \cc spectroscopy has followed the discovery
of the remarkable X$(3872)$ 
by Belle \cite{Choi:2003ue} and CDF \cite{Acosta:2003zx} 
in B decays to $J/\psi \pi^+ \pi^-$; 
assuming that this is a real resonance rather than a threshold effect,
the X$(3872)$ is presumably either a DD$^*$ charmed meson molecule 
\cite{Close:2003sg,Swanson:2003tb,Tornqvist:2004qy}
or a narrow $\J = 2$ D-wave \cc state \cite{Barnes:2003vb,Eichten:2004uh}.
Very recent observations of the X$(3872)$ in
$\gamma J/\psi$ and $\omega J/\psi$ by Belle 
support a $1^{++}$ DD$^*$ molecule
assignment \cite{Abe:2005ix,Abe:2005iy}.

There has also been experimental activity in the spin-singlet 
P-wave sector, with recent reports of the observation of the elusive 
$1^1{\rm P}_1$ $h_c$ state by CLEO \cite{Seth:2005pr,Tomaradze:2004sk}.
Finally, the surprisingly large cross sections
for double charmonium production in $e^+e^-$ 
reported by Belle
\cite{Abe:2002rb,Abe:2004ww,Pakhlov:2004au,PakPri} 
suggest that it may be possible to study C $= (+)$ $c\bar c$ states
in $e^+e^-$ without using the higher-order  
$O(\alpha^4)$ two-photon annihilation process.

One open topic of great current interest 
in \cc spectroscopy is the search for the 
$\psi_2(1^3{\rm D}_2)$ and $\eta_{c2}(1^1{\rm D}_2)$ states, 
which are expected to be quite narrow due to the absence of 
open-charm decay modes. 

A second topic is the Lorentz nature of confinement; in pure \cc models 
this is tested by the multiplet splittings of orbitally-excited \cc states.
For example, with pure scalar confinement as is normally assumed 
there is no spin-spin hyperfine interaction at O$(v^2/c^2)$, so the masses of 
spin-singlets (such as the $^1{\rm P}_1$ $h_c$) 
are degenerate with the corresponding triplet c.o.g. (center of gravity);
here this is the $^3$P$_{\rm J}$ c.o.g., at 3525~MeV. 
In the original Cornell model 
\cite{Eichten:1978tg}
it was assumed that confinement acts as 
the time component of a Lorentz vector, which lifts the degeneracy
of the $h_c$ and the $^3$P$_{\rm J}$ c.o.g. 
Another possibility is that confinement may
be a more complicated mix of scalar and timelike vector 
\cite{Ebert:2005jj}.
Of course these simple potential model 
considerations may be complicated by mass shifts due to 
other effects, such as couplings to open-flavor channels
\cite{Eichten:2004uh}. 

A third topic is the search for exotica such as hybrids; 
the level of mixing between conventional quarkonium and hybrid basis states 
falls rapidly with increasing quark mass, which 
suggests that nonexotic hybrids may be more easily distinguished from 
conventional quarkonia in charmonium than in the light quark sectors.
Since LGT (lattice gauge theory) predicts that the lightest \cc hybrids lie near 4.4~GeV 
\cite{Bernard:1997ib,Liao:2002rj,Mei:2002ip,Bali:2003tp},
there is a strong incentive to establish the ``background" spectrum 
of conventional \cc states up to and somewhat beyond this mass. 

A final topic of current interest is the importance of mixing between
quark model $q\bar q$ basis states and two-meson continua, which has been 
cited as a possible reason for the low masses of the recently discovered
D$_{s\J}$ states \cite{Aubert:2003fg,Besson:2003jp}.
The effects of ``unquenching the quark model" by
including meson loops can presumably be studied effectively in the 
\cc system, in which the experimental spectrum of states 
is relatively unambiguous. 
The success of the $q\bar q$ quark model is surprising, in
view of the probable importance of corrections to the valence
approximation; the range of validity of the naive ``quenched" $q\bar q$
quark model is an interesting and open question 
\cite{Swanson:2003ec}.

Motivated by this revived interest in \cc spectroscopy, we have carried 
out a theoretical study of the expected properties of charmonium states, 
notably the poorly understood higher-mass \cc levels above DD threshold.
Two variants of potential models are used in this study, a conventional
nonrelativistic model based on the Schr\"odinger equation with a 
Coulomb plus linear potential, and the Godfrey-Isgur relativized 
potential model. 
We give results 
for all states in the multiplets 1-4S, 1-3P, 1-2D, 1-2F and 1G,
comprising 40 \cc resonances in total.
Predictions are given for quantities which are likely to be of 
greatest experimental 
interest, which are the spectrum of states, 
E1 (and some M1) electromagnetic transition rates, and
strong partial and total widths of states above open-charm threshold. 

Similar results for
many of the electromagnetic transition rates have recently been 
reported by Ebert {\it et al.} \cite{Ebert:2002pp}.
The $\ell^+\ell^-$ leptonic and two-photon 
widths are not discussed in detail here, 
as they have been considered extensively elsewhere; see for example
\cite{Barnes:2004cz} 
and
\cite{Li:1990sx,Ackleh:1991ws,Ackleh:1991dy}
and references cited therein.

\section{Spectrum}

\subsection{Nonrelativistic potential model}

As a minimal model of the charmonium system we use a nonrelativistic
potential model, with wavefunctions determined by the 
Schr\"odinger equation with a conventional quarkonium potential.
We use the standard color Coulomb plus linear scalar form, and also include 
a Gaussian-smeared contact hyperfine interaction in the zeroth-order potential.
The central potential is 
\be
V_0^{(c\bar c)}(r) = -\frac{4}{3}\frac{\alpha_s}{r} + br
+ 
\frac{32\pi\alpha_s}{9 m_c^2}\, 
\tilde \delta_{\sigma}(r)\,
\vec {\S}_c \cdot \vec {\S}_{\bar c}\,
\ee  
where
$\tilde \delta_{\sigma}(r) = (\sigma/\sqrt{\pi})^3\, e^{-\sigma^2 r^2}$. 
The four parameters $(\alpha_s, b, m_c, \sigma)$  are determined 
by fitting the spectrum.

The spin-spin contact hyperfine interaction is one of the spin-dependent 
terms predicted by one gluon exchange (OGE) forces. The contact form 
$\propto \delta(\vec x\,)$ is actually an artifact of an O$(v_q^2/c^2)$ 
expansion 
of the T-matrix \cite{Barnes:1982eg}, so replacing it by an interaction with 
a range $1/\sigma$ comparable to $1/m_c$ is not an unwarranted modification.

We treat the remaining spin-dependent terms as mass shifts using 
leading-order perturbation theory. These are the OGE spin-orbit 
and tensor and a
longer-ranged inverted spin-orbit term, which arises from the assumed
Lorentz scalar confinement. These are explicitly
\be
V_{spin-dep} = 
\frac{1}{m_c^2}\, \bigg[
\Big(
\frac{2\alpha_s}{r^3}  
-
\frac{b}{2 r} 
\Big)
\, \vec {\L} \cdot \vec {\S}  
+
\frac{4\alpha_s}{r^3} \, \T \ 
\bigg] .
\ee

The spin-orbit operator is diagonal in a $|\J,\L,\S\rangle$ basis, with 
the matrix elements 
$\langle \vec {\L} \cdot \vec {\S} \rangle = 
[\J(\J+1) - (\L(\L+1) - \S(\S+1)]/2$.
The tensor operator T has nonvanishing diagonal matrix elements only between
$\L > 0$ spin-triplet states, which are 

\be
\langle ^3{\rm L}_{\J} | \T | ^3{\rm L}_{\J} \rangle = 
\cases{
   -\frac{\L}{6(2\L + 3)}, & $\J = \L + 1$ \cr
& \cr
   +\frac{1}{6},  & $\J = \L$ \cr
& \cr
   -\frac{(\L + 1)}{6(2\L - 1)}, & $\J = \L-1$ \cr
}
\ .
\ee
For experimental input we use the masses of the 11 reasonably
well established $c\bar c$ states, which are given in Table~\ref{Table_spectrum}
(rounded to 1~MeV). The parameters that follow from fitting these masses
are $(\alpha_s, b, m_c, \sigma) = (0.5461,  0.1425~{\rm GeV}^2,  
1.4794~{\rm GeV},  1.0946~{\rm GeV})$. Given these values, we can 
predict the masses and matrix elements of the currently unknown \ccbar 
states; Table~\ref{Table_spectrum} and Fig.~\ref{spectrum} show the predicted 
spectrum.

\subsection{Godfrey-Isgur relativized potential model}

The Godfrey-Isgur model is a ``relativized'' extension of the nonrelativistic 
model of the previous section. This model assumes
a relativistic dispersion relation for the quark kinetic energy, 
a QCD-motivated running coupling $\alpha_s(r)$, a flavor-dependent 
potential smearing parameter $\sigma$, and replaces 
factors of quark mass with quark kinetic energy. Details of the model 
and the method of solution may be found in Ref.\cite{Godfrey:1985xj}. 
The Hamiltonian consists of a relativistic kinetic term and 
a generalized quark-antiquark potential,
\be
{\rm H} = {\rm H}_0 + \V_{q\bar{q}} (\vec{p},\vec{r})
\ee
where
\be
{\rm H}_0 = 
\sqrt{\vec p_q^{\, 2} +m_q^2} 
+ \sqrt{\vec p_{\bar q}^{\, 2} + m_{\bar{q}}^2 }\ .
\ee
Just as in the nonrelativistic model, the quark-antiquark potential 
V$_{q\bar{q}} (\vec{p},\vec{r})$ assumed here incorporates the
Lorentz vector one gluon exchange interaction at
short distances and a Lorentz scalar linear confining interaction.
To first order in $(v_q/c)^2$, V$_{q\bar{q}} (\vec{p},\vec{r}\, )$
reduces to the standard non-relativistic result given by Eqns.(1) and
(2) (with $\alpha_s$ replaced by a running coupling constant, $\alpha_s(r)$).
The full set of model parameters is given in Ref.\cite{Godfrey:1985xj}.
Note that the string tension and quark mass ($b =  0.18$ GeV$^2$ and
$m_c=1.628$ GeV) are significantly larger than the values used in 
our nonrelativistic model. 

One important aspect of this model is that it gives reasonably accurate 
results for the spectrum and matrix elements of
quarkonia of all $u,d,s,c,b$ quark flavors, whereas the 
nonrelativistic model of the previous section is only fitted to
the \cc system.

\subsection{Discussion}

\begin{figure}[h]
\includegraphics[width=8.5cm,angle=0]{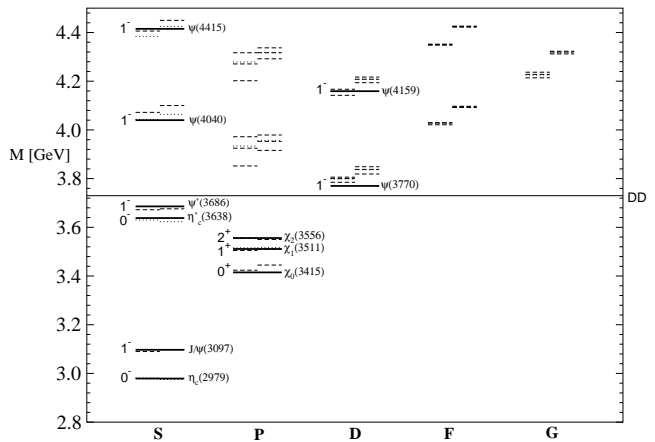}
\caption{
Predicted and observed spectrum of charmonium states
(Table~\ref{Table_spectrum}). The
solid lines are experiment, and the 
broken lines are theory (NR model left, 
GI right). Spin triplet levels are dashed, and spin singlets
are dotted. The DD open-charm threshold at 3.73 GeV is also shown.}
\label{spectrum} 
\end{figure}

The spectra predicted by the NR and GI models 
(Table~\ref{Table_spectrum} and Fig.~\ref{spectrum}) 
are quite similar for S- and P-wave states, largely because of the 
constraints provided by the experimental \cc candidates for these 
multiplets.
We note in passing that these potential model results are very similar
to the most recent predictions of the charmonium spectrum from LGT
\cite{Okamoto:2001jb,Liao:2002rj,Hashimoto:2004fv}.
At higher L we have only the L=2 $1^3$D$_1$ and $2^3$D$_1$ states 
$\psi(3770)$ and $\psi(4159)$
to constrain the models, and the predicted
mean D-wave multiplet masses differ by {\it ca.} 50~MeV. 
For $\L > 2$ the absence of experimental 
states allows a relatively large scatter of predicted mean masses, 
which differ by as much as $\approx 100$~MeV in the 1G multiplet. 
(The splittings within higher L multiplets in contrast are 
rather similar.) The mean multiplet masses predicted 
by the two models differ largely because of the values assumed for the
string tension $b$, 
which is $0.18$~GeV$^2$ in the GI model but is
a rather smaller $0.1425$~GeV$^2$ in the NR model. 
Identification of any L=3 or L=4 \cc state would be very useful 
as a constraint on the spectrum of higher-L \cc states generally. 
This is unfortunately a difficult problem, since these states 
are not easily accessible experimentally. As we shall see, one possibility 
is to produce the $^3$F$_2$ $\ \chi_2$ \cc state, which is formed in E1 
radiative transitions from the $\psi(4159)$ 
and decays dominantly to
DD and DD$^*$.
(For simplicity in this paper we abbreviate the final state D$\bar {\rm D}$ 
as ``DD", the state D$\bar {\rm D}^* + \bar {\rm D}$D$^*$ 
as ``DD$^*$", and so forth.)

\vskip 1.0cm
\section{Radiative transitions}
 
\subsection{E1 transitions}

Radiative transitions of higher-mass charmonium states are of interest
largely because they provide one of the few pathways between 
\cc states with different quantum numbers. Since typical E1 radiative partial 
widths of charmonia are 10s to 100s of keV, corresponding 
to significant branching fractions of $\sim 10^{-3}$ to $10^{-2}$, 
large event samples of radially excited S-wave states produced 
in $e^+e^-$ annihilation could be used to identify 
radially excited P-wave states, which are not otherwise
easily produced. Similarly, the E1 radiative transition of the nominally 
2$^3$D$_1$ $\psi(4159)$ can be used to produce an F-wave \cc state; 
this multiplet would likely be very difficult to reach using other mechanisms.

We evaluate these E1 radiative partial widths using 
\begin{widetext}
\be
\Gamma_{\rm E1}(
{\rm n}\, {}^{2{\S}+1}{\rm L}_{\J} 
\to 
{\rm n}'\, {}^{2{\S}'+1}{\rm L}'_{{\J}'}  
+ \gamma) 
 =  \frac{4}{3}\, 
C_{fi}\,
\delta_{{\S}{\S}'} \, 
e_c^2 \, \alpha \,
|\,\langle \psi_f
|
\, r \,
|\, 
\psi_i
\rangle\, |^2  
\,
{\rm E}_{\gamma}^3 \, 
\frac{\E_f^{(c\bar c)}}{\M_i^{(c\bar c)}}\ ,  
\ee
\end{widetext}
where $e_c= 2/3$ is the $c$-quark charge in units of $|e|$,
$\alpha$ is the fine-structure constant,
E$_{\gamma}$ is the final photon energy, 
$\E_f^{(c\bar c)}$ is the total energy of the final \cc state, 
$\M_i^{(c\bar c)}$ is the mass of the initial \cc state,   
the spatial matrix element 
$\,\langle \psi_f|\, r \, |\,\psi_i\rangle$ 
involves the initial and final radial wavefunctions,
and the angular matrix element
$C_{fi}$ is 
\be 
C_{fi}=\hbox{max}({\L},\; {\L}')\; (2{\J}' + 1)
\left\{ { {{\L}' \atop {\J}} {{\J}' \atop {\L}} {{\S} \atop 1}  } \right\}^2 .
\ee
This is the result of Ref.\cite{Kwong:1988ae}, 
except for our inclusion of 
the relativistic phase space factor of $\E_f^{(c\bar c)}/\M_i^{(c\bar c)}$, 
which is usually not far from unity. 
(The GI results do not include this phase space factor.)
We evaluate these radiative partial widths in both the NR and GI  
models. For the NR model the matrix elements 
$\langle {n'}^{2{\S}'+1}{\L}'_{{\J}'} |\; r \; 
| n^{2{\S}+1}{\L}_{\J}  \rangle$ 
were evaluated using the Coulomb plus linear plus smeared hyperfine 
wavefunctions of the potential model described in Sec.IIA, and for the  
GI model they were evaluated 
using the wavefunctions of Ref.\cite{Godfrey:1985xj}. Since the 
masses predicted for unknown states differ in the two
models, the assumed photon energy E$_{\gamma}$ differs  
as well; these photon energies are given in the 
E1 and M1 transition tables
(Tables~\ref{E1rad_StoP}-\ref{M1rad})
together with the radiative partial widths.

Some E1 transitions that are of special importance for the study 
of higher charmonium states are discussed in the text. Transitions 
from initial $1^{--}$ \cc states are of greatest interest in this
regard, since these can be studied with high statistics at $e^+e^-$
machines. These can provide access to the spin-triplet members of 
the 2P and 3P multiplets in particular, starting from the $\psi(4040)$ 
and $\psi(4415)$. E1 radiative transitions may also be useful in 
identifying the narrow 1D $3^-$ and $2^-$ \cc states, since they are 
all predicted to have large partial widths ({\it ca.} 300~keV)
to the 1P $\chi_{\J}$ and $h_c$ states. 

\subsection{M1 transitions}

Although M1 rates are typically rather weaker than E1 rates, 
they are nonetheless interesting because they may allow access 
to spin-singlet states that are very difficult to produce otherwise. 
It is also interesting that the known M1 rates
show serious disagreement between theory and experiment. 
This is in part due to the fact that M1 transitions between different 
spatial multiplets, such as $\psi' \to \gamma \eta_c$ 
(2S $\to$ 1S), are nonzero only due to small relativistic corrections 
to a vanishing lowest-order M1 matrix element.  

The M1 radiative partial widths are evaluated using 

\begin{widetext}
\be
\Gamma_{\rm M1}(
{\rm n}\, {}^{2{\S}+1}{\rm L}_{\J} 
\to 
{\rm n}'\, {}^{2{\S}'+1}{\rm L}'_{{\J}'}  
+ \gamma) 
 =  
\frac{4}{3}\,  
{2J'+1\over 2L+1}\,
\delta_{{\L}{\L}'} \, 
\delta_{{\S},{\S}'\pm 1} \, 
e_c^2 \, 
\frac{\alpha}{m_c^2}\,
|\,\langle \psi_f
|\, 
\psi_i
\rangle\, |^2  
\,
{\rm E}_{\gamma}^3 \, 
\frac{\E_f^{(c\bar c)}}{\M_i^{(c\bar c)}}\,  
\ .
\label{M1rate}
\ee
\end{widetext}
(See the previous E1 formula for definitions.) 
The GI M1 radiative rates do not incorporate the 
$\E_f^{(c\bar c)}/\M_i^{(c\bar c)}$ phase space factor, 
but do include a $j_0(kr/2)$ recoil factor. 
We quote NR results both with and without this recoil factor.
The photon energies E$_{\gamma}$ depend on the model
in most cases, since we have assumed theoretical masses for unknown states. 

As with the E1 rates, we quote M1 
radiative widths in both the NR potential model of Sec.IIA and the 
GI model \cite{Godfrey:1985xj} of Sec.IIB.
Although the M1 decay rates involve a relatively simple off-diagonal
matrix element of the magnetic moment operator, we note that our
predicted rates show considerable variation with the model assumptions,
and agreement with the two known M1 rates is not good. In the off-diagonal
(2S $\to$ 1S) transition $\psi' \to \gamma \eta_c$ 
this is because the zeroth-order
M1 matrix element vanishes due to orthogonality of the spatial wavefunctions,
and the nonzero predicted rate results from rather model-dependent corrections
to the wavefunctions and recoil effects. Evidently the corrections we have
included do not accurately predict the observed partial width; for this 
reason the other predicted M1 rates between different spatial multiplets
are suspect, and in any case are evidently strongly dependent on recoil
factors. Better experimental data will be very important for improving 
our description of these apparently simple but evidently poorly understood 
M1 radiative transitions. 
\section{Open-flavor strong decays}

\vskip -0.2cm
\subsection{The Decay Model}

\vskip -0.2cm
The dominant strong decays (when allowed by phase space) are 
transitions to open-flavor final states. In these open-flavor decays
the initial \cc meson decays through production of a light 
$q\bar q$ quark-antiquark pair $(q=u,d,s)$, followed
by separation into two open-charm mesons. Remarkably, the QCD mechanism 
underlying this
dominant decay process is still poorly understood. 
In quark model calculations this decay process is 
modelled by a simple phenomenological \qq pair production 
amplitude. The \qq pair is usually assumed to be produced with 
vacuum $(0^{++})$ quantum numbers, and
variants of the decay model make different 
assumptions regarding
the spatial dependence of the pair production amplitude 
relative to the initial \cc pair. 
The simplest of these models is the $^3$P$_0$ model, 
originally introduced by Micu~\cite{Micu:1968mk}, which
assumes that the new $q\bar q$ pair is produced with vacuum 
quantum numbers ($^3$P$_0$) by a spatially constant 
pair-production amplitude $\gamma$.

LeYaouanc~{\it et al.} subsequently applied 
the $^3$P$_0$ model to meson \cite{LeYaouanc:1972ae}
and baryon \cite{LeYaouanc:1973xz,LeYaouanc:1978ef}
open flavor strong decays in a series of publications in the 1970s.
They also evaluated 
strong decay partial widths
of the three \ccbar states 
$\psi(3770)$, 
$\psi(4040)$ 
and  
$\psi(4415)$
in the $^3$P$_0$ model 
\cite{LeYaouanc:1977ux,LeYaouanc:1977gm}; 
the relation between this
early work and our contribution will be discussed.
 
This decay model, which has since been applied extensively to the 
decays of light mesons and baryons, was originally adopted 
largely because of the success in predicting the D/S amplitude ratio
in the decay $b_1\to \omega \pi$ 
\cite{LeYaouanc:1972ae,Kokoski:1985is,Ackleh:1996yt,Nozar:2002br}.
Another stringent test of strong decay models is provided by
the very tight limit on the spin singlet to spin singlets transition 
$\pi_2(1670) \to b_1\pi$ from the VES Collaboration  
\cite{Amelin:1998qr,PDG},
\begin{equation} 
B_{\pi_2 (1670)\to b_1\pi}\ < \; 1.9 \cdot 10^{-3}, \ \ 97.7\% \ \ c.l. 
\end{equation}
This branching fraction is predicted to be zero in the $^3$P$_0$ model, 
but would not necessarily be negligible in a different decay model 
or if final state interactions are important. 
(Final state interactions combined with the 
$^3$P$_0$ model allow two-stage transitions such as 
$\pi_2 \to \rho\pi \to b_1\pi$, although the direct 
decay 
$\pi_2 \to b_1\pi$ is forbidden in the $^3$P$_0$ model.) 

Recent variants of the $^3$P$_0$ model modify the 
pair production vertex~\cite{Roberts:1997kq} or modulate the spatial 
dependence of the pair production amplitude to simulate 
a gluonic flux tube \cite{Kokoski:1985is}.
(The latter is the ``flux-tube decay model", which 
gives very similar predictions to the $^3$P$_0$ model in practice.) 

Another class of decay models assumes that the pair production amplitude
transforms as the time component of a Lorentz vector. 
The Cornell group used a decay model
of this type for charmonium \cite{Eichten:1978tg}. 
This model appears to describe the partial 
and total widths of some charmonium states reasonably well, 
although it is known to disagree with experimental 
amplitude ratios in the light quark sector
\cite{Ackleh:1996yt}.

In this work we employ a formalism that is equivalent to
the original constant-amplitude version of the $^3$P$_0$ decay model,
although we have simplified the calculations
by deriving momentum-space Feynman rules \cite{Ackleh:1996yt} 
instead of using the real-space convolution integrals 
of LeYaouanc~{\it et al.} \cite{LeYaouanc:1972ae}.
  
In our formalism the $^3$P$_0$ model describes decay matrix elements
using a \qq pair production Hamiltonian which is the nonrelativistic 
limit of 
\be
H_{decay} = \gamma\, \sum_q 2 m_q \int d^{\, 3} x\; \ \bar \psi_q \psi_q \ .
\ee
Here $\psi_q$ is a Dirac quark field, $m_q$ is the constituent quark mass, 
and $\gamma$ is the dimensionless \qq pair production amplitude, which is fitted
to data. The decay arises from the matrix element of this model Hamiltonian
between an initial meson state $|{\rm A}\rangle$ and a final meson 
pair $|{\rm BC}\rangle$, which is nonzero due to the 
$b^{\dagger}d^{\dagger}$ term. Here we will use SHO 
(simple harmonic oscillator) 
wavefunctions for the mesons, 
with a universal width parameter $\beta$.
The pair-production strength parameter 
$\gamma$ is found to be roughly flavor-independent in light meson decays 
involving pair production of $u\bar u$, $d\bar d$ and $s\bar s$ pairs. 
In our recent extensive studies of light ($u,d,s$) meson decays
\cite{Ackleh:1996yt,Barnes:1996ff,Barnes:2002mu} 
we found that the value 
\be
\gamma = 0.4
\ee 
gives a reasonably accurate description of the 
overall scale of decay widths.   

\subsection{Charmonium Strong Decays}

\subsubsection{Previous studies}

The open-flavor decay amplitudes of the three \ccbar states 
$\psi(3770)$, $\psi(4040)$ and $\psi(4415)$ were evaluated
in the $^3$P$_0$ model by LeYaouanc~{\it et al.} in the late 1970s 
\cite{LeYaouanc:1977ux,LeYaouanc:1977gm}. Their calculations are generally
quite similar to the approach used here, although we note
three important differences:

\vskip 0.2cm
\noindent
1) LeYaouanc~{\it et al.} 
fitted the pair production strength $\gamma$ 
and wavefunction length scale $R$ to the
charmonium decay data, whereas we used the same pair production amplitude 
$\gamma$ as in light meson decays, and our SHO wavefunction length scale $\beta^{-1}$ 
is taken from \ccbar potential models.

\vskip 0.2cm
\noindent
2) We use the correct reduced mass coordinates for the unsymmetric
open-charm meson wavefunctions, whereas LeYaouanc~{\it et al.}
assumed symmetric SHO wavefunctions for all mesons. 
This approximation proves to be quite important numerically,
since $m_c \gg m_s, m_{u,d}$. 

\vskip 0.2cm
\noindent
3) We consider all energetically allowed open-flavor decay modes. 
LeYaouanc~{\it et al.} did not consider some S+P decays of the 
$\psi(4415)$, nor did they consider the important
J$^{\P} = 1^+$ D meson singlet-triplet mixing angle. 
(The masses of the P-wave charmed mesons were unknown
at that time.) 

\vskip 0.2cm
Although we have improved on the earlier calculations 
of LeYaouanc~{\it et al.}, and evaluate a much more extensive set 
of decay amplitudes, we do concur with their important conclusion that 
the DD mode of the $\psi(4040)$ is naturally suppressed given a 
$3^3$S$_1$ \ccbar assignment, and also that S+P modes are important 
for the $\psi(4415)$.

Other more recent applications of the $^3$P$_0$ decay model to charmonium 
include a study of the strong decays of various $c\bar c$ X(3872) 
candidates to DD \cite{Barnes:2003vb}, and an
examination of the hypothesis that
the $\psi(4040)$ and $\psi(4160)$ are linear superpositions of a \cc
hybrid and a conventional $\psi({\rm 3S})$ \cc state \cite{Close:1995eu}.

The only other detailed studies of the open-flavor strong decays of charmonia
of which we are aware are the early work by the Cornell group 
\cite{Eichten:1978tg},
and recent more detailed applications of this model to other 
charmonium states \cite{Eichten:2004uh}.
The decay model of the Cornell group assumes a pair creation operator 
which is the time component of a Lorentz vector ($j^0$), rather than the 
Lorentz scalar assumed by the $^3$P$_0$ model. As we noted previously,
this $j^0$ model does not agree well with some light meson decay 
amplitude ratios. In the original Cornell model decay calculations 
only the combined channel contributions to $R$ were derived, 
and only channels containing pairs of S-wave open-charm mesons were included. 
The very interesting relative amplitudes in the D$^*$D$^*$ channel have not 
been evaluated in this model.

\subsubsection{This study}

Tables~\ref{strong_3S4S}-\ref{strong_1G}
present the partial widths and strong decay amplitudes
we find for all kinematically allowed open-flavor decay 
modes of all the charmonium states listed in 
Table~\ref{Table_spectrum}.
We have evaluated these in the
$^3$P$_0$ model, assuming SHO wavefunctions with a width parameter of 
$\beta = 0.5$~GeV and a pair production amplitude of $\gamma = 0.4$. 
SHO wavefunctions have the advantage that the decay amplitudes 
can be determined analytically,
and we have found that the numerical results are usually not 
strongly dependent on the 
details of the spatial wavefunctions 
\cite{Kokoski:1985is,Ackleh:1996yt,Geiger:1994kr,Blundell:1995ev},
unless they are near a node. 
This width parameter was chosen by comparing the overlap of 
NR and GI potential model 
wavefunctions with SHO wavefunctions, which was found to be largest 
for both charmonium 
and open-charm mesons for a $\beta$ near 0.5~GeV.     

\begin{figure}[h]
\includegraphics[width=6cm,angle=-90]{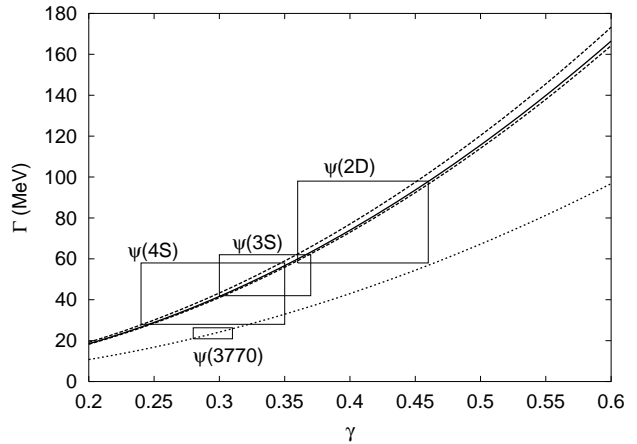}
\caption{
Values of the $^3$P$_0$ decay model pair production strength $\gamma$
implied by the experimental total widths of the known higher-mass
vector states $\psi(3770)$, $\psi(4040)$, $\psi(4159)$ and $\psi(4415)$.
This calculation assumes pure \cc spectroscopic states, respectively
$1^3\D_1$, $3^3\S_1$, $2^3\D_1$ and $4^3\S_1$. 
The boxes show current PDG experimental width uncertainties.}
\label{gamma} 
\end{figure}

As a test of the accuracy of these parameters, especially the 
value assumed for the \qq pair production amplitude $\gamma$ 
(which is taken from light meson decays), 
in Fig.~\ref{gamma} we compare the predicted total widths
for $\beta = 0.5$~GeV and variable $\gamma$
to the values observed for the four known \cc states above 
open-charm threshold, $\psi(3770)$, $\psi(4040)$, $\psi(4159)$ 
and $\psi(4415)$.
Although there is some scatter in the value of $\gamma$ specified
by these widths, a choice of $\gamma = 0.35$ is evidently near optimum, 
and yields an average error of only 29\%, 
which is very reasonable for this phenomenological model.
In the decay tables we will quote numerical results for our standard 
light-meson value of $\gamma = 0.4$, which is rather close to
this value. A subsequent change in $\gamma$ will simply scale the 
widths as $\gamma^2$.

For this paper the strong decay amplitudes were first evaluated analytically, 
following which numerical
values were determined given our parameters $\beta = 0.5$~GeV,  
$\gamma = 0.4$, 
and our constituent quark masses 
$m_{u,d} = 0.33$~GeV,
$m_s     = 0.55$~GeV
and
$m_c = 1.5$~GeV.
(Quark mass {\it ratios} are required to specify the final 
open-charm meson wavefunctions.) We determine the decay kinematics
using external meson masses taken from the Particle Data Group 
\cite{PDG} (charge averaged where appropriate)
and from recent Belle results \cite{Abe:2003zm}.
These masses are 
M$_{\D}                  = 1.867$ GeV, 
M$_{\D^*}                = 2.008$ GeV, 
M$_{\D_0^*}              = 2.308$ GeV \cite{Abe:2003zm}, 
M$_{\D_1}$(narrow)$      = 2.425$ GeV,  
M$_{\D_1'}$(broad)$      = 2.427$ GeV \cite{Abe:2003zm},
M$_{\D_2^*}              = 2.459$ GeV,
M$_{\D_s}                = 1.968$ GeV,
M$_{\D_s^*}              = 2.112$ GeV, 
M$_{\D_{s0}^*}           = 2.317$ GeV, 
M$_{\D_{s1}}             = 2.459$ GeV.
The FOCUS collaboration \cite{Vaandering:2004ix}
estimate a somewhat higher mass for the scalar $\D_0^*$, however 
their result is complicated by the presence of 
both scalar and broad axial vector ($\D_1'$) contributions
\cite{Kut04}.

The $\J^{\P} = 1^+$ axial vector $c\bar n$ and $c\bar s$ mesons
$\D_1$ and $\D_1'$ are assumed to be coherent superpositions
of quark model spin-singlet and spin-triplet states, 

\begin{eqnarray}
&|\D_1\rangle  &= 
+\cos(\theta) |^1\P_1\rangle + 
\sin(\theta) | ^3\P_1\rangle 
\nonumber 
\\
&|\D_1'\rangle  &= 
-\sin(\theta) |^1\P_1\rangle + 
\cos(\theta) | ^3\P_1\rangle \ .
\end{eqnarray}
We define this singlet-triplet mixing angle in the LS coupling scheme, 
in accord with the conventions of Ref.\cite{Barnes:2002mu}. 
This reference also discusses other conventions for this 
mixing angle that have appeared in the literature.
In the heavy-quark limit we expect to find a ``magic" mixing angle, 
due to the quark mass dependence of the spin-orbit and
tensor terms, which is $\theta = 35.3^o (-54.7^o)$
if the expectation of the heavy-quark spin-orbit interaction 
is negative (positive) \cite{Godfrey:1986wj}.
In the following we assume the first case, which is supported by the 
reported widths of the $1^+$ P-wave $c\bar n$ mesons.
We note however that finite quark mass effects and mixing induced 
by higher-order Fock states can substantially modify this mixing angle, 
so it should more generally be treated as a free parameter. 
We will discuss the dependence of our strong decay amplitudes on this 
mixing angle for a few sensitive cases.

\section{Discussion of \cc states}

\subsection{Known states above 3.73~GeV}

The four known \cc states above DD threshold, 
$\psi(3770)$,
$\psi(4040)$,
$\psi(4159)$ and
$\psi(4415)$
are of special interest because they are easily produced at $e^+e^-$ machines.
Accordingly we will first discuss our predicted strong decay amplitudes and
widths for these states. This will be followed by a more general
discussion of \cc strong decays by multiplet. 

\subsubsection{$\psi(3770)$}

The $\psi(3770)$ is generally assumed to be the $^3$D$_1$ \cc state, 
perhaps with a significant $2^3$S$_1$ component
\cite{Rosner:2004wy,Rosner:2004mi}.
(This additional component can explain the leptonic width,
which is much larger than expected for a pure $^3$D$_1$ \cc state
\cite{Barnes:2004cz}.)

\begin{figure}[h]
\includegraphics[width=8cm]{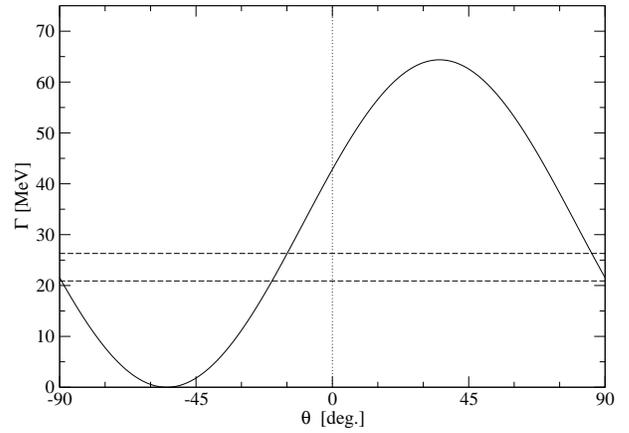}
\caption{
The total width of the $\psi(3770)$ as a function of the 
$^3$D$_1$ - 2$^3$S$_1$ mixing angle $\theta$. The experimental 
$1\sigma$ limits
are shown as horizontal dashed lines.}
\label{3770_mixing} 
\end{figure}

For a pure $^3$D$_1$ state at the $\psi(3770)$ mass we predict a 
DD width of 43~MeV with our parameters, which is rather larger than 
the experimental value of $23.6 \pm 2.7$~MeV. The partial width
of a mixed 2S-D state
\be
|\psi(3770)\rangle =
+ 
\cos(\theta) |{}^3{\rm D}_1\rangle
+
\sin(\theta) |2{}^3{\rm S}_1\rangle
\ee
is shown in Fig.~\ref{3770_mixing}
as a function of the mixing angle $\theta$. 
In this simple model, fitting the experimental 
$\psi(3770)$ width requires a mixing angle of 
$\theta = -17.4^o$~$\pm$~$2.5^o$. Assuming that the leptonic widths scale 
as the S-wave component squared, this mixing angle predicts an $e^+e^-$
partial width ratio of
\be
\frac{\Gamma_{e^+e^-}(\psi(3770))}
{\Gamma_{e^+e^-}(\psi(3686))}\bigg|_{thy.} 
= 0.10 \pm 0.03 \ .
\ee

It is interesting that this is consistent with experiment,
\be
\frac{\Gamma_{e^+e^-}(\psi(3770))}
{\Gamma_{e^+e^-}(\psi(3686))}\bigg|_{expt.} 
= 0.12 \pm 0.02 \ ,
\ee
so the small $\psi(3770)$ total width may indeed be an effect of
$^3$D$_1$ - 2$^3$S$_1$ mixing.

\vskip 1cm

\subsubsection{$\psi(4040)$}

Next in mass is the $\psi(4040)$, which is a very interesting case 
for the study of strong decays. Four open-charm modes are
energetically allowed,
DD, 
DD$^*$,
D$^*$D$^*$
and
D$_s$D$_s$,
although D$^*$D$^*$ has little phase space. 
There is some experimental evidence for the three nonstrange modes
from Mark~I at SLAC \cite{Goldhaber:1977qn}.
Remarkably, the reported relative branching fractions (scaled by $p^3$)
show a very strong preference for D$^*$ final states, 
$\D^{*0}\D^{*0} >> \D^0\D^{*0} >> \D^0\D^0$.
This motivated suggestions that the $\psi(4040)$ might be a 
$\D^*\D^*$ molecule \cite{Voloshin:1976ap,DeRujula:1976qd,Iwao:1980iw}.
The Mark~I results were
\bd
\frac{\B}{p^3} 
(\; \D^{*0}\D^{*0} \ \; : \; \ \ \D^0\D^{*0} \ \ : \ \  \D^0\D^0 \; ) \ =\
\hskip 2cm
\ed
\vskip -0.8cm
\be
128\pm 40 \; : \; 4.0 \pm 0.8 \; : \; 0.2\pm 0.1 \ .
\hskip 1.0cm
\ee
This is much less striking once the $p^3$ factors
are restored, which gives
\bd
\B
(\; \D^{*0}\D^{*0} \ \, : \ \ \ \ \D^0\D^{*0} \ \ \ \, : \ \  \D^0\D^0 \; ) \ =\
\hskip 2cm
\ed
\vskip -0.8cm
\be
1\pm 0.31 \; : \; 0.95 \pm 0.19 \; : \; 0.12\pm 0.06 \ ,
\hskip 0.8cm
\ee
where we have used the $\D^{*0}\D^{*0}$ mode to set the scale.
These relative branching fractions should be compared to our theoretical 
$\psi(4040)$ branching fractions, not the more often quoted $\B/p^3$ ratios.

Comparison with Table~\ref{strong_3S4S} 
shows that the predicted $^3$P$_0$ model 
partial and total widths for a 3$^3$S$_1$ $\psi(4040)$ state 
are actually in accord with 
the Mark~I results. The total width is predicted to be 74~MeV,
which is reasonably close to the PDG experimental average of $52\pm 10$~MeV.
Seth \cite{Seth:2004py} estimates a rather larger width of $88\pm 5$~MeV 
for this state from Crystal Ball and new BES data, 
which is closer to our predicted width.

The $\D\D^*$ and $\D^*\D^*$ branching fractions are predicted to 
be approximately equal, consistent with Mark~I. 
The~DD mode is predicted to be much smaller,
despite its larger phase space, 
as it is accidentally near a node in the $^3$P$_0$ model decay amplitude.
(This was previously noted by
LeYaouanc {\it et al.} \cite{LeYaouanc:1977ux}.)
This set of predictions constitutes a very nontrivial success of the 
$^3$P$_0$ decay model. The actual width to the suppressed DD mode is probably 
a few MeV; the 0.1~MeV quoted in the table depends strongly on the location 
of the node, and varying the SHO width parameter $\beta$ by $\pm 10\%$ suggests
that a DD partial width of a few MeV is plausible. This is again consistent 
with the Mark~I result.

The $\D^*\D^*$ mode is especially interesting for strong decay studies, 
since there are three independent decay amplitudes for 
$1^{--}\; c\bar c \to \D^*\D^*$, 
$^1$P$_1$,
$^5$P$_1$ and
$^5$F$_1$. If the $\psi(4040)$ is a pure S-wave \cc state, 
we predict a
zero $^5$F$_1$ $\D^*\D^*$ decay amplitude, and the ratio of the nonzero
amplitudes is
$^5$P$_1 /  ^1$P$_1 = -2\sqrt{5}$ 
(independent of the 
radial wavefunction). As we shall see when we discuss the $\psi(4159)$,
very different $\D^*\D^*$ amplitude ratios are predicted for a
$^3$D$_1$ initial \cc state.

The unobserved mode D$_s$D$_s$ is predicted to have a branching fraction of 
about $11\%$; this prediction is fortunately not especially 
sensitive to the location of the decay amplitude node for 
$\psi(4040)$ to two pseudoscalars. 
This branching fraction is of special 
interest because it determines the event rates available for studies of 
$\D_s$ weak decays at $e^+e^-$ colliders such as CLEO.

Finally, we note that a large sample of $\psi(4040)$ events might be
used to access the 2P \cc multiplet, since the E1 radiative branching
fractions for $\psi(4040) \to \gamma \chi'_{\J}$ are expected to be 
{\it ca.} $10^{-3}$ (Table~\ref{E1rad_StoP}). 

\subsubsection{$\psi(4159)$}

Comparison of the mass of the $1^{--}$ $\psi(4159)$ 
with potential model predictions (Fig.~\ref{spectrum}) immediately 
suggests a 2$^3$D$_1$ \cc assignment. 
There may also be a significant
S-wave \cc component, since the $\psi(4159)$ has a much larger
$e^+e^-$ width than one would expect for a pure D-wave \cc
state \cite{Barnes:2004cz}. 
(This was also the case for the $\psi(3770)$.)

There are five open-flavor decay modes available to a $1^{--}$ \cc 
vector at this mass,
DD, 
DD$^*$,
D$^*$D$^*$,
D$_s$D$_s$
and
D$_s^{\phantom{*}}$D$_s^*$. 
The predicted decay amplitudes and partial widths for a 2$^3$D$_1$ 
$\psi(4159)$ 
are given in Table~\ref{strong_1D2D}. The theoretical total width of 
74~MeV is in good agreement with the experimental value of
$78\pm 20$~MeV. Seth \cite{Seth:2004py} quotes a similar width of 
$107\pm 8$~MeV for this state from Crystal Ball and recent BES data.

The leading mode is predicted to be D$^*$D$^*$
with a branching fraction of $\approx 50\%$, followed by comparable 
DD and D$_s^{\phantom{*}}$D$_s^*$ modes (both $\approx 20\%$), 
and a somewhat weaker D$_s$D$_s$ ($\approx 10\%$). The DD$^*$ 
branching fraction is predicted to be very small, since 
it is suppressed by a decay amplitude node near the physical point.

The mode D$^*$D$^*$ is again especially interesting, due to the 
three decay amplitudes allowed for this final state. 
(The other modes have only single amplitudes.) For a pure D-wave \cc
assignment the ratio of the two $\D^*\D^*$ 
P-wave amplitudes is independent
of the radial wavefunction, and is
$^5$P$_1 /  ^1$P$_1 = -1/\sqrt{5}$. The $^5$F$_1$ amplitude is predicted to
be the largest given this 2$^3$D$_1$ assignment, whereas it is zero for an
S-wave \cc state. Clearly, a determination of these D$^*$D$^*$ decay amplitude
ratios would be an extremely interesting test of decay models.  
Experimentally, to date nothing has been reported regarding the 
exclusive hadronic decay modes of the $\psi(4159)$.

\subsubsection{$\psi(4415)$}

The final \cc resonance known above DD threshold is the 
$1^{--}$ $\psi(4415)$.
Again, the mass of this resonance relative to potential model predictions
suggests a \cc assignment, in this case 4$^3$S$_1$. Of course this 
identification requires independent confirmation, since \cc hybrids are 
predicted to first appear near this mass by LGT simulations 
\cite{Bernard:1997ib,Liao:2002rj,Mei:2002ip,Bali:2003tp},
and the lightest hybrid multiplet in the flux tube model includes a
$1^{--}$ state \cite{Isgur:1985vy}.

Ten open-charm strong decay modes are allowed for 
the $\psi(4415)$, seven with 
$c\bar n$ meson final states ($n=u,d$), and three with $c\bar s$. 
The predicted branching fractions of a 4$^3$S$_1$ \cc state
at this mass are quite characteristic, 
and (if the $^3$P$_0$ decay model is accurate) 
may be useful in confirming this assignment.
As a note of caution, a 4S state has three radial nodes, 
and some of the smaller predicted branching fractions are sensitive to the 
locations of the nodes.

The predicted total width of a 4$^3$S$_1$ \cc meson at this mass 
with our parameters is 77~MeV, somewhat larger than
the experimental PDG average width of $43\pm 15$~MeV. Seth \cite{Seth:2004py} 
notes however that the width of this state in Crystal Ball and recent
BES data is rather larger, and quotes an average of $119\pm 15$~MeV, 
on the opposite side of the $^3$P$_0$ decay model prediction. 

The largest exclusive mode is predicted to be the 
S+P combination DD$_1$, where D$_1$ is the narrower of the two 
$1^+$ $c\bar n$ axial mesons near 2.42-2.43~GeV. Since the D$_1$ is rather narrow 
($\Gamma \approx 20$-$30$~MeV) and decays dominantly to D$^*\pi$, 
there should be a strong $\psi(4415)$ signal in DD$^*\pi$ final states. 
Although both S-wave and D-wave DD$_1$ final states are allowed, 
in the $^3$P$_0$ model the HQET (heavy quark effective theory) 
D$_1$ mixing angle $\theta$ 
is just the value needed to give a zero S-wave  
$\psi(4415) \to \D{\D}_1$ amplitude. Thus
we have the striking prediction that the dominant 
$\psi(4415)$ decay mode is DD$_1$, in D-wave rather than S-wave.

The second-largest $\psi(4415)$ branching fraction is predicted to be another 
S+P mode, DD$_2^*$. The D$_2^*$ is also moderately narrow,
so this isobar can also be isolated 
from the observed final state. 
(The PDG quotes a neutral D$_2^*$ total width of
$\Gamma = 23 \pm 5 $~MeV, but Belle \cite{Abe:2003zm} and
FOCUS \cite{Link:2003bd} 
find somewhat broader values of 
$\Gamma = 45.6 \pm 8.0 $~MeV and 
$\Gamma = 38.7 \pm 5.3 \pm 2.9 $~MeV respectively; 
see \cite{Coo04} for a recent discussion.
The D$_2^*$ has significant branching fractions 
to both D$^*\pi$ and D$\pi$, so the DD$_2^*$ mode of the $\psi(4415)$ 
should be observable in both DD$\pi$ and DD$^*\pi$.    

A final important $\psi(4415)$ mode is predicted to be D$^*$D$^*$, 
which should be comparable in strength to DD$_2^*$. As noted in the previous 
discussions of $\psi(4040)$ and $\psi(4159)$ decays, D$^*$D$^*$ is 
an especially interesting decay mode because it has three amplitudes, 
and the $^5$P$_1$/$^1$P$_1$ amplitude ratio is independent of the radial
wavefunction for pure S-wave or D-wave \cc states. If the 
$\psi(4415)$ is indeed an S-wave (4$^3$S$_1$) \cc state to a 
good approximation, we expect this ratio to be 
$^5$P$_1$/$^1$P$_1 = -2\sqrt{5}$, and the $^5$F$_1$ amplitude 
should be zero. 

\begin{figure}[ht]
\includegraphics[width=8.5cm]{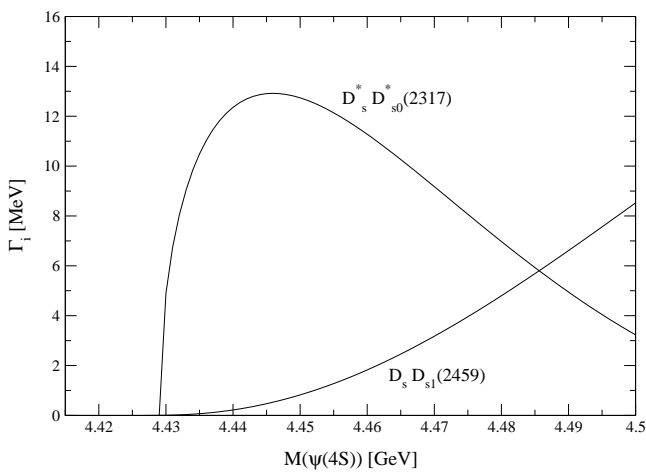}
\caption{
Partial widths predicted for $\psi(4^3\S_1) \to \D^*_s \D^*_{s0}(2317)$ and
$\D_s \D_{s1}(2459)$ as a function of the assumed $\psi(4^3\S_1)$ mass. 
}
\label{4415_scan}
\end{figure}

It is interesting to note that $\psi(4415)$ decays may provide 
access to the recently discovered D$_{s0}^{*}(2317)$. 
Although the channel D$_s^{*}$D$_{s0}^*(2317)$ has a threshold of 
4429~MeV, 14~MeV above the nominal mass of the $\psi(4415)$,
the width of the $\psi(4415)$ and the fact that 
the decay $\psi(4415) \to $D$_s^{*}$D$_{s0}^*(2317)$ is purely S-wave
implies that one may observe 
significant D$_{s0}^*(2317)$ production 
just above threshold, near E$_{cm} \approx 4435$~MeV. 
This is illustrated in Fig.~\ref{4415_scan}, in which we show 
theoretical $^3\P_0$ decay model partial widths to
D$^*_s$D$^*_{s0}(2317)$ 
and
D$_s$D$_{s1}(2459)$
as functions of the 
$4^3$S$_1$ \ccbar mass. (This calculation assumes pure 
$c\bar s$ D$^*_{s0}(2317)$ and D$_{s1}(2459)$ states, 
and should be modified accordingly if they have significant 
non-$c\bar s$ components.)
Unfortunately, $\psi(4415)$ decays are not expected to be 
similarly effective in producing the D$_{s1}(2459)$, 
because the assumed ``magic-mixed" HQET state
$|\D_{s1}(2459)\rangle = 
\sqrt{1/3}\; |^3\P_1(c\bar s)\rangle
+
\sqrt{2/3}\; |^1\P_1(c\bar s)\rangle 
$ 
predicts a vanishing
$\psi(4415) \to \D_s \D_{s1}$ 
S-wave decay amplitude. 

\vskip 1cm

\subsection{3S and 4S states}

The two unknown states in the 3S and 4S multiplets are the
3$^1$S$_0$ 
and
4$^1$S$_0$
pseudoscalars. To evaluate their strong decays we have assigned 
potential model masses to these states, specifically 4043~MeV and 4384~MeV. 
The resulting total widths are predicted to be moderate, 
80~MeV and 61~MeV respectively, and both states should be observable 
in DD$^*$ and D$^*$D$^*$ 
(see Table~\ref{strong_3S4S}).
No other open-charm 
strong decay modes are allowed for a 3$^1$S$_0$ $\eta_c(4043)$.
Although these modes are also important for the 4$^1$S$_0$ $\eta_c(4384)$,
its largest branching fraction is predicted to be to the S+P combination
DD$_2^*$ ($\approx 40\%$).

These states may be observable 
in $\gamma\gamma$ and hadronic production (through $gg$ fusion), 
analogous to the $\eta_c$ and $\eta_c'$. 
M1 transitions from the higher vectors to these states 
are unfortunately predicted to be rather weak, since they are
either hindered or have little phase space. 
(See Table~\ref{M1rad}.) 

\subsection{2P states}
  
The mean 2P multiplet mass is predicted to be near 3.95~GeV.
There is an interesting disagreement between the NR and GI models 
regarding the scale 
of mass splittings within this multiplet, as well as in the 3P states. 
(See Table~\ref{Table_spectrum} and Fig.~\ref{spectrum}.) 

There are few open-flavor strong modes available to the 2P states 
(Table~\ref{strong_2P3P}).
One result of this restricted phase space 
is the prediction of a fairly narrow 2$^3$P$_0$ $\chi_0(3852)$ 
(assuming the NR model mass) that decays only to DD, with a total width of
just 30~MeV. This small width is due in part to a decay amplitude node, 
and a variation of the total width by a factor of two could easily be 
accommodated. The two axial states 
2$^3$P$_1$ $\chi_1(3925)$
and
2$^1$P$_1$ $h_c(3934)$
can only decay to DD$^*$, and both have nonzero S- and D-wave
decay amplitudes; their D/S ratios are $-0.18$ and $+0.46$ respectively.
Observation of the D/S ratio in a DD$^*$ enhancement could be a useful 
check of the assumption of a resonant $1^+$ \cc contribution.    

The relatively narrow 2$^3$P$_0$ state 
$\chi_0(3852)$ can be produced in $\gamma\gamma$, 
as can its somewhat wider 2$^3$P$_2$ $\chi_2(3972)$ partner. 
Additional possible production mechanisms include $gg$ fusion
(for the 2$^3$P$_0$ and 2$^3$P$_2$), E1 radiative 
transitions from higher $1^{--}$ \cc states (to all 2P $\chi_{\J}$ states),
and B decays. 
The E1 branching fractions of the $\psi(4040)$ and $\psi(4159)$ to 
2P $\chi_{\J}$ states are predicted to be sufficiently large 
({\it ca.}~$10^{-3}$, see Tables~\ref{E1rad_StoP},\ref{E1rad_1D2D})
to allow identification of these states at BES and CLEO. 

Although no 2P $c\bar c$ state has been clearly established 
experimentally, there are two recent reports of enhancements 
which may be due to states 
in this multiplet. The Belle Collaboration has reported 
evidence of an enhancement in $\omega J/\psi$ with a mass and width of 
$3941 \pm 11$~MeV and $92\pm 24$~MeV \cite{Abe:2004zs}; this
is compatible with
expectations for 2P $\chi_{\J}$ $c\bar c$ states.
(Although 2P $c\bar c$ states should strongly favor open-charm 
strong decay modes, inelastic FSIs (final state interactions)
will allow weaker transitions
to closed-charm final states such as $\omega J/\psi$. These FSIs 
should be most important in S-wave, as in
the virtual second-order processes
$\chi_{0,2}(3940) \to \D^*\D^* \to \omega J/\psi$.) 
A search for evidence of this state in open-charm final states,
with much larger branching fractions than to $\omega J/\psi$, 
will be a crucial test of a charmonium assignment.  
Assuming that the enhancement is indeed due to 
a 2P $\chi_{\J}$ state, if it is the $0^{++}$ state it will populate only DD,
if $1^{++}$ only DD$^*$, and if $2^{++}$ comparable branching fractions
to DD and DD$^*$ are expected.

There is a second report of a possible 2P state, 
also from the Belle Collaboration \cite{Pakhlov:2004au},
in double charmonium production. 
An enhancement is seen in $e^+e^- \to {\rm J}/\psi\; \D\D^*$ 
at a mass of $3940 \pm 12$~MeV, with a limit on the total width  
of $< 96$~MeV ($90\% \ c.l.$). As this enhancement is not seen
in $\omega J/\psi$, it has been suggested \cite{PakPri} 
that it is distinct from the
3940~MeV signal discussed above. Consideration of possible 2P candidates 
(Table~\ref{strong_2P3P}) suggests that this enhancement may be due to the
$\chi_{c1}(2$P$)$, which is predicted to have a similar mass of 3925~MeV. 
(The predicted total width however is a somewhat larger 165~MeV.) 

\subsection{3P states}

The 3P states have an expected mean multiplet mass of about 4.3~GeV. 
As with the 2P states there is a significant difference 
between the NR and GI models
in the scale of mass splittings predicted within the multiplet.
This leads for example to a 90~MeV difference between the models
for the predicted mass of the 3$^3$P$_0$ scalar.

Many open-charm strong decay channels are open at the 3P mass scale.
The individual decay amplitudes however are typically 
somewhat smaller than for the 2P states, due in part to nodes in the 
decay amplitudes. This has the interesting result that the predicted 
mean total width of 3P states is smaller than for 2P states  
(58~MeV for 3P versus 91~MeV for 2P).

The predicted branching fraction to the mode D$^*$D$^*$ 
is large for all four 3P states,
and is the largest for all except the spin-singlet
3$^1$P$_1$ $h_c(4279)$; in this case the leading mode is DD$_0^*$.
For all 3P states except the 3$^3$P$_1$
the D$^*$D$^*$ decay mode has multiple amplitudes, which 
are predicted to be comparable in strength. 
An experimental determination of any of these amplitude ratios would 
provide an interesting test of decay models. The decays
of the tensor 3$^3$P$_2$ $\chi_2(4317)$ to DD$_1$ and DD$_1'$ 
are also interesting, in that the prediction that the broad D$_1'$ 
mode dominates is a test of the axial-D mixing angle.   

As with the 2P states, the most important experimental problem is to
find an effective production mechanism. 
Again, $\gamma\gamma$ and $gg$ fusion can produce the 
$2^{++}$ and $0^{++}$ states (here 3$^3$P$_2$ and 3$^3$P$_0$), 
although there should be moderate suppression of the production amplitudes 
relative to the 1P ground states due to the higher masses and smaller 
short-distance \cc wavefunctions.    
E1 transitions from the $\psi(4415)$ could lead to the identification of the 
3P states, since radiative partial widths of 50-100~keV are expected 
to the 3$^3$P$_2$ and 3$^3$P$_1$; these correspond to branching fractions 
of {\it ca.}~$10^{-3}$. Production of the 3$^3$P$_0$ state however 
is expected to 
be suppressed, due to a radiative decay amplitude node. 

\subsection{1D states}

Since the $^3$D$_2$ and $^1$D$_2$ states do not have 
allowed open-flavor decay modes, and we have already considered
$^3$D$_1$ decays in the section on the $\psi(3770)$,
we restrict our discussion to the 
$^3$D$_3$. This currently unknown state is predicted
to have a mass of 3806~MeV in the NR potential model.
It is expected to be quite narrow simply because it has 
little phase space to the only open-flavor mode DD, which
has an F-wave centrifugal barrier; with our parameters 
the predicted total strong width is just 0.5~MeV. 
The radiative width for the E1 transition 
$\psi_3(3806) \to \gamma \chi_2$ is predicted to be about 0.3~MeV 
(Table~\ref{E1rad_1D2D}), so the total width 
of this $^3$D$_3$ state should be near 1~MeV, 
with comparable strong and radiative branching fractions. 

Finding an effective production mechanism for this interesting state 
is again a crucial problem. As it has C~$=(-)$, it cannot be made in 
$\gamma \gamma$ or $gg$ fusion. One possibility is that the $^3$D$_3$ 
may be observable in E1 transitions from the radial $2^3$P$_2$ tensor, 
since the radiative partial width is predicted to be a relatively large 
$\approx 100$~keV. Similarly, the narrow $^3$D$_2$ and $^1$D$_2$ states
may be observable in E1 transitions from 2P states, assuming that a production 
mechanism can be found for 2P states. All three of these states can 
also be produced in B-decays \cite{Eichten:2002qv}.

\subsection{2D states}

The states in the 2D multiplet are predicted to be essentially
degenerate in both the NR and GI potential models.
The $2^3$D$_1$ candidate $\psi(4159)$ suggests
a 2D multiplet mass near 4.16~GeV, which is consistent with 
potential model expectations. Since the $\psi(4159)$ was discussed
previously, here we will consider only the unknown states
$2^3$D$_3$, $2^3$D$_2$ and $2^1$D$_2$. 

These states are predicted to be rather broad, 
with theoretical total widths ranging from 78~MeV for the 
$2^3$D$_1$ (consistent with the $\psi(4159)$)
to 148~MeV for the $2^3$D$_3$.  
The energetically allowed open-flavor modes are all of S+S type,
since the first S+P mode is DD$^*_0$, at a nominal mass of 4175~MeV.
The decays of 2D states are dominated by DD, DD$^*$ and D$^*$D$^*$, 
with smaller contributions from the charm-strange 
final states D$_s$D$_s$ and D$_s^{\phantom{*}}$D$_s^*$. 

If these states (besides the $\psi(4159)$) can be produced,
there are several interesting tests of their strong decay amplitudes.
The $2^3$D$_3$ $\psi_3(4167)$, for example, is predicted to have D$^*$D$^*$ 
as the dominant mode, with both $^1$F$_3$ and $^5$F$_3$ amplitudes present,
in the ratio $^5$F$_3$/$^1$F$_3 = -\sqrt{24/5}$. The remaining 2D states
have many multiamplitude decays, in which the higher-L partial waves have 
large or dominant amplitudes. (Compare the P- and F-waves in the decays
in Table~\ref{strong_1D2D}.)

\subsection{1F states}

The four states in the 1F multiplet are predicted to be almost 
degenerate in both the NR and GI models, although the models disagree
somewhat regarding this mass. In the NR potential model 
the states are expected near 4025~MeV, whereas in the GI model the 
expected mass is near 4095~MeV. This discrepancy is mainly due to the 
different values assumed for the string tension. 

It is unfortunate that production amplitudes for these higher-L states 
are expected
to be quite weak, since the 1F multiplet is predicted to 
contain a rather narrow 
state, the $^3$F$_4$ $\chi_4$ (Table~\ref{strong_1F}). Assuming our NR model 
$^3$F$_4$ mass of 4021~MeV, we predict a strong width for this state of just 
8.3~MeV. This is partly a result of the centrifugal barrier (both open modes, 
DD and DD$^*$, are G-waves), and is also because there is essentially 
no phase space for D$^*$D$^*$. If the GI masses are more accurate we would 
no longer expect a very narrow $^3$F$_4$ state, since the D$^*$D$^*$ 
partial width increases rapidly above threshold. At the GI mass of 4095~MeV 
the $^3$F$_4$ is predicted to have a partial width to D$^*$D$^*$ of 38~MeV, 
and a total width of 60~MeV.

The single well-established mechanism that might be exploited to
produce a 1F state is an E1 radiative transition from the $\psi(4159)$
to the $^3$F$_2$ $\chi_2$.  This transition however has a rather weak, 
model-dependent partial width of 20-50~keV (Table~\ref{E1rad_1D2D}), and
the $^3$F$_2$ is predicted to be the broadest 1F state, with
a total width of about 160~MeV.

If an effective production mechanism for 1F states is identified,
the E1 decays of these states can be used to 
populate the narrow 1D states; the E1 transitions 
$^3$F$_4 \to {}^3$D$_3$,
$^3$F$_3 \to {}^3$D$_2$
and
$^1$F$_3 \to {}^1$D$_2$
are predicted to have rather large radiative widths of {\it ca.}~300-400~keV
(see Table~\ref{E1rad_1F}).

\subsection{2F and 1G states}

To complete our study we have evaluated the spectrum and decays of 
all states in the 2F and 1G multiplets
(Tables~\ref{strong_2Fa} and \ref{strong_1G}), 
although the lack of a 
clear experimental route to these states suggests that it may be 
quite difficult to test these predictions.

The multiplets are again predicted to be nearly degenerate in both the 
NR and GI potential models, with somewhat lower 
masses predicted by the NR model. The mean 2F and 1G masses are 
approximately 4350~MeV and 4225~MeV respectively in the NR model, 
and 4425~MeV and 4320~MeV in the GI model.

Assuming NR model masses,
these states are predicted to have total widths that range from 
a minimum of 58~MeV for the $^3$G$_5$ $\psi_5(4214)$
to a maximum of 180~MeV for the $2^3$F$_2$ $\chi_2(4351)$. 
The individual
decay modes do show some interesting features; although the dominant modes
are usually of S+S type, some S+P modes are important in 2F decays. 
For example, the largest decay modes of the 
$2^3$F$_3$ $\chi_3(4352)$
and
$2^3$F$_2$ $\chi_2(4351)$  
are both S+P, with branching fractions of 
B($\chi_3 \to $ DD$^*_2) = 30\% $ 
and
B($\chi_2 \to $ DD$_1) = 59\% $ 
respectively.

\section{Summary and Conclusions}

We have computed the spectrum,
all allowed E1 and some M1 electromagnetic partial widths, 
and all allowed open-charm strong decay amplitudes
of the 40 charmonium states expected to {\it ca.} 4.4~GeV. 
These were evaluated using two potential models, 
the quark model formulation of electromagnetic 
transition amplitudes, and the $^3\P_0$ strong decay model.

The predictions of the spectrum should be useful in the identification 
of new states, and in tests of the Lorentz structure of 
confinement and the nature of spin-dependent forces. 
The transition amplitudes will be useful in searches for 
new higher-mass \cc states, as they suggest which final states should
be populated preferentially by the decays of a given initial \cc state, 
as well as predicting the often characteristic strong decay amplitude
ratios within a given final state. These predictions may also be useful 
in distinguishing between conventional charmonia and exotica 
such as charmed meson molecules (perhaps including the X(3872)) 
and \cc hybrids. These results all implicitly test the accuracy of the 
pure-\cc assumption made in these models, and can serve as benchmarks 
for calculations that relax this assumption. 

Future experimental measurements of 
strong partial widths and decay amplitude ratios
can provide very important tests of strong decay models, 
the $^3$P$_0$ model in particular. We have discussed many 
specific examples of these tests in the previous sections; 
here we will summarize some of our most important observations. 

The $\psi(3770)$ is the lightest known \cc state above open-charm threshold. 
We noted that the relatively large $e^+e^-$ width of the $\psi(3770)$ 
is difficult to understand if it is assumed to be a pure $^3$D$_1$ \cc state,
but this problem is solved if the $\psi(3770)$ 
has a significant admixture of the $2^3\S_1$ $c\bar c$ basis state. 
This mixing also solves the $\psi(3770)$ total width problem,
since the mixing angle required to fit the $e^+e^-$ width 
is consistent with the value required to fit the $^3$P$_0$ model 
DD strong width to experiment. This mixing angle 
has been discussed in earlier references
(see for example Refs.\cite{Rosner:2004wy,Rosner:2004mi}), 
although the observation that the
same mixing angle resolves both the $e^+e^-$ and total width discrepancies
has not been noted previously. 

The relative branching fractions of the $\psi(4040)$  to 
DD, DD$^*$ and D$^*$D$^*$, although often cited as anomalous,
are naturally explained in the $^3\P_0$ model by a node
in the DD decay amplitude. A very important test of this 
strong decay model and the nature of the $\psi(4040)$ follows 
from a simple measurement of the ratios of the three independent 
D$^*$D$^*$ decay amplitudes. The $^3\P_0$ model predicts 
that the $^5\F_1$ amplitude is zero, and the ratio of the two 
P-wave amplitudes is $^5\P_1/^1\P_1 = -2 \sqrt{5}$, independent of the 
radial wavefunction, provided that the $\psi(4040)$
is a pure S-wave \cc state. Similarly, the D$^*$D$^*$ amplitude ratios 
in $\psi(4159)$ decays test the decay model and the 
assumed $2^3\D_1$ assignment for this state. The $^5\F_1$  
D$^*$D$^*$ amplitude is predicted to be the largest for a pure $2^3\D_1$ 
$\psi(4159)$, and the P-wave amplitude ratio
is predicted to be $^5\P_1/^1\P_1 = -1 /\sqrt{5}$, again independent 
of the radial wavefunction. Finally, we noted that 
it may be possible to identify a higher-L \cc state 
(the $^3\F_2$ $\chi_2$ member of the F-wave \cc multiplet) 
in E1 radiative $\psi(4159)$ decays.

The $\psi(4415)$ has ten open-flavor modes, and their branching fractions 
have never been determined. We predict that the largest is the unusual 
S+P combination DD$_1$. It is especially notable that strong decays of 
the $\psi(4415)$ may provide a novel production mechanism for the
enigmatic $\D_{s0}^*(2317)$, through the above-threshold decay 
$\psi(4415+\epsilon ) \to \D_s^* \D_{s0}^*(2317)$, as shown in 
Fig.~\ref{4415_scan}. One may also produce higher-mass charmonium states 
from the $\psi(4415)$ through electromagnetic transitions.
For example, $\psi(4415)$ E1 decays are predicted to produce 3P states 
with branching fractions at the per mil level. These states could then be 
identified through their subsequent open-charm strong decays. 

As discussed in Section V.C, two new states near 3940~MeV have recently been 
reported by Belle. 2P $\chi_{\J}$ states are natural assignments, 
as they have theoretical quark model masses ranging
from 3852 to 3972 MeV and widths of 50-100 MeV, similar to the reported
values. A~measurement of the relative branching fractions to  
$\D\D$, $\D\D^*$ and $\D^*\D^*$ and comparisons to our $^3$P$_0$
decay model predictions should allow the determination of the quantum numbers
of these states, and will show whether they are indeed consistent with 
2P $c\bar c$ assignments.

A better determination of the properties of conventional charmonium states 
at higher masses is important not only because of the improved understanding
of QCD that will follow (especially aspects of confinement and strong decays), 
but also because qualitatively different types of resonances are expected
at these masses. 
These new states include charmed meson 
molecules and charmonium hybrids, and the identification
of these novel excitations will obviously be easier if the conventional 
charmonium spectrum is well established. 

The recent results from B factories and new programs at   
BES, CLEO and GSI have led to a resurgence of interest in the physics of
charmonium. We argue that a detailed experimental investigation of the
spectrum of excited charmonium states and their decay properties 
will considerably improve our understanding of the nonperturbative
aspects of QCD.

\acknowledgments

We acknowledge useful discussions with
R.Galik, T.Pedlar, C.Quigg, J.Rosner, K.Seth and T.Swarnicki 
in the course of this work. 

This research was supported in part by 
the Natural Sciences and Engineering Research Council of Canada, 
the U.S. National Science
Foundation through grant NSF-PHY-0244786 at the University of Tennessee, 
and the U.S. Department of Energy under contracts 
DE-AC05-00OR22725 at Oak Ridge National Laboratory and 
DE-FG02-00ER41135 at the University of Pittsburgh, 
and by PPARC grant PP/B500607.

\begin{table*}
\caption{Experimental and theoretical spectrum of \ccbar states.
The experimental masses are PDG averages, 
which are rounded to 1~MeV and assigned
equal weights in the theoretical fits.
For the $2^1{\rm S}_0$ $\eta_c'(3638)$ we use
a world average of recent measurements \cite{Swa03}.} 
\vskip 0.5cm
\begin{tabular}{cr|lc|cc} 
\hline
Multiplet\ \  & State\phantom{,,} 
&\qquad \phantom{,,}Expt.\phantom{,,} 
&\phantom{,,}Input (NR)\phantom{,,} 
& \multicolumn{2}{c}{Theor.\qquad }\\
& & & &\phantom{,,,}\quad NR\quad \phantom{,,,}
&\phantom{,,,}\quad GI \quad \phantom{,,,}\\ 
\hline
1S &  $J/\psi(1^3{\rm S}_1) $ & \quad $ 3096.87 \pm 0.04 $ & 3097 & 3090 & 3098\\
   &  $\eta_c(1^1{\rm S}_0) $ & \quad $ 2979.2 \pm 1.3   $ & 2979 & 2982 & 2975\\
\hline
2S &  $\psi'(2^3{\rm S}_1) $ & \quad $ 3685.96 \pm 0.09 $ & 3686 & 3672 & 3676\\
   &  $\eta_c'(2^1{\rm S}_0) $ & \quad $ 3637.7 \pm 4.4   $ & 3638 & 3630 & 3623\\
\hline
3S &  $\psi(3^3{\rm S}_1) $ & \quad $ 4040 \pm 10      $ & 4040 & 4072 & 4100\\
   &  $\eta_c(3^1{\rm S}_0) $ &                      &      & 4043 & 4064\\
\hline
4S &  $\psi(4^3{\rm S}_1) $ & \quad $ 4415 \pm 6       $ & 4415 & 4406 & 4450 \\
   &  $\eta_c(4^1{\rm S}_0) $ &                      &      & 4384 & 4425 \\
\hline
1P &  $\chi_2(1^3{\rm P}_2) $ & \quad $ 3556.18 \pm 0.13 $ & 3556 & 3556 & 3550\\
   &  $\chi_1(1^3{\rm P}_1) $ & \quad $ 3510.51 \pm 0.12 $ & 3511 & 3505 & 3510\\
   &  $\chi_0(1^3{\rm P}_0) $ & \quad $ 3415.3 \pm 0.4   $ & 3415 & 3424 & 3445\\
   &  $h_c(1^1{\rm P}_1) $    & \qquad see text            &      & 3516 & 3517\\
\hline
2P &  $\chi_2(2^3{\rm P}_2) $ &                      &      & 3972 & 3979\\
   &  $\chi_1(2^3{\rm P}_1) $ &                      &      & 3925 & 3953\\
   &  $\chi_0(2^3{\rm P}_0) $ &                      &      & 3852 & 3916\\
   &  $h_c(2^1{\rm P}_1) $ &                         &      & 3934 & 3956\\
\hline
3P &  $\chi_2(3^3{\rm P}_2) $ &                      &      & 4317 & 4337\\
   &  $\chi_1(3^3{\rm P}_1) $ &                      &      & 4271 & 4317\\
   &  $\chi_0(3^3{\rm P}_0) $ &                      &      & 4202 & 4292\\
   &  $h_c(3^1{\rm P}_1) $ &                         &      & 4279 & 4318\\
\hline
1D &  $\psi_3(1^3{\rm D}_3) $ &                      &      & 3806 & 3849\\
   &  $\psi_2(1^3{\rm D}_2) $ &                      &      & 3800 & 3838\\
   &  $\psi(1^3{\rm D}_1) $ & \quad  $ 3769.9 \pm 2.5  $ 
                                                     & 3770 & 3785 & 3819\\
   &  $\eta_{c2}(1^1{\rm D}_2) $ &                   &      & 3799 & 3837\\
\hline
2D &  $\psi_3(2^3{\rm D}_3) $ &                      &      & 4167 & 4217\\
   &  $\psi_2(2^3{\rm D}_2) $ &                      &      & 4158 & 4208\\
   &  $\psi(2^3{\rm D}_1) $ & \quad  $ 4159 \pm 20     $ 
                                                     & 4159 & 4142 & 4194\\
   &  $\eta_{c2}(2^1{\rm D}_2) $ &                   &      & 4158 & 4208\\
\hline
1F &  $\chi_4(1^3{\rm F}_4) $ &                      &      & 4021 & 4095\\
   &  $\chi_3(1^3{\rm F}_3) $ &                      &      & 4029 & 4097\\
   &  $\chi_2(1^3{\rm F}_2) $ &                      &      & 4029 & 4092\\
   &  $h_{c3}(1^1{\rm F}_3) $ &                      &      & 4026 & 4094\\
\hline
2F &  $\chi_4(2^3{\rm F}_4) $ &                      &      & 4348 & 4425\\
   &  $\chi_3(2^3{\rm F}_3) $ &                      &      & 4352 & 4426\\
   &  $\chi_2(2^3{\rm F}_2) $ &                      &      & 4351 & 4422\\
   &  $h_{c3}(2^1{\rm F}_3) $ &                      &      & 4350 & 4424\\
\hline
1G &  $\psi_5(1^3{\rm G}_5) $ &                      &      & 4214 & 4312\\
   &  $\psi_4(1^3{\rm G}_4) $ &                      &      & 4228 & 4320\\
   &  $\psi_3(1^3{\rm G}_3) $ &                      &      & 4237 & 4323\\
   &  $\eta_{c4}(1^1{\rm G}_4) $ &                      &      & 4225 & 4317\\
\hline
\hline
\end{tabular}
\label{Table_spectrum}
\end{table*}

\begin{table*}
\caption{S $\to$ P E1 radiative transitions in the NR
and GI potential models.
The masses are taken from Table~\ref{Table_spectrum}; we use the  
experimental masses (rounded ``input" column) 
if known, and for the $^1$P$_1$ $h_c$ we assume a mass of 3525~MeV,
which is the c.o.g. of the $^3$P$_{\rm J}$ $\chi_{\J}$ states. 
Otherwise, theoretical values are used.}
\vskip 0.3cm
\begin{ruledtabular}
\begin{tabular}{l r r c c c c c c } 
Multiplets 
& Initial meson  & Final meson 
& \multicolumn{2}{c}{E$_{\gamma}$ (MeV)} 
& \multicolumn{2}{c}{$\Gamma_{\rm thy}$~(keV)} 
& $\Gamma_{\rm expt}$~(keV) & \\
&                &             & NR  & GI  & NR & GI &   \\
\hline
\\
2S $\to$ 1P  
& $\psi'(2^3{\rm S}_1) $ & $\chi_{2}(1^3{\rm P}_2)$  & 128.  & 128.  &  38. 
& 24.  & 27. $\pm$ 4.   \\
\\
&                        & $\chi_{1}(1^3{\rm P}_1)$  & 171.  & 171.  &  54. 
& 29.  & 27. $\pm$ 3.   \\
\\
&                        & $\chi_{0}(1^3{\rm P}_0)$  & 261.  & 261.  &  63. 
& 26.  & 27. $\pm$ 3.   \\
\\
&$\eta_c'(2^1{\rm S}_0)$ & $h_c(1^1{\rm P}_1)$ 
                                                    & 111.  & 119.  &  49. & 36.  & \\
\\
\hline
\\
3S $\to$ 2P  
& $\psi(3^3{\rm S}_1) $  
& $\chi_{2}(2^3{\rm P}_2)$  
& 67.  &  119. &  14. & 48.  & \\ 
\\
&                        
& $\chi_{1}(2^3{\rm P}_1)$  
& 113.  & 145.  & 39. & 43.  & \\
\\
&                        
& $\chi_{0}(2^3{\rm P}_0)$  
& 184.  & 180.  & 54. &  22. & \\
\\
& $\eta_c(3^1{\rm S}_0) $ & $h_c(2^1{\rm P}_1)$ 
& 108.  & 108.  &  105. & 64.  & \\
\\
\hline
\\
3S $\to$ 1P  
& $\psi(3^3{\rm S}_1) $  & $\chi_{2}(1^3{\rm P}_2)$  & 455.  & 508.  &  0.70 & 12.7  & \\
\\
&                        & $\chi_{1}(1^3{\rm P}_1)$  & 494.  & 547.  &  0.53 & 0.85  & \\
\\
&                        & $\chi_{0}(1^3{\rm P}_0)$  & 577.  & 628.  &  0.27 & 0.63  & \\
\\
& $\eta_c(3^1{\rm S}_0) $ & $h_c(1^1{\rm P}_1)$ 
                                                    & 485.  & 511.  &  9.1 &  28. & \\
\\
\hline
\\
4S $\to$ 3P  
& $\psi(4^3{\rm S}_1) $  
& $\chi_{2}(3^3{\rm P}_2)$  
& 97.  & 112. &   68. &   66.  & \\ 
\\
&                        
& $\chi_{1}(3^3{\rm P}_1)$  
& 142.  & 131.  & 126. &   54.  & \\
\\
&                        
& $\chi_{0}(3^3{\rm P}_0)$  
& 208.  & 155.  & 0.003 &   25. & \\
\\
& $\eta_c(4^1{\rm S}_0) $ 
& $h_c(3^1{\rm P}_1)$ 
& 104.  & 106.  &  159. & 101.  & \\
\\
\hline
\\
4S $\to$ 2P  
& $\psi(4^3{\rm S}_1) $  
& $\chi_{2}(2^3{\rm P}_2)$  
& 421.  & 446. &   0.62 &   15.  & \\ 
\\
&                        
& $\chi_{1}(2^3{\rm P}_1)$  
& 423.  & 469.  &  0.49 &   0.92  & \\
\\
&                        
& $\chi_{0}(2^3{\rm P}_0)$  
& 527.  & 502.  &  0.24 &  0.39 & \\
\\
& $\eta_c(4^1{\rm S}_0) $ 
& $h_c(2^1{\rm P}_1)$ 
& 427.  & 444.  & 10.1 &  31.3  & \\
\\
\hline
\\
4S $\to$ 1P  
& $\psi(4^3{\rm S}_1) $  
& $\chi_{2}(^3{\rm P}_2)$  
& 775.  & 804. &  0.61  &  5.2  & \\ 
\\
&                        
& $\chi_{1}(^3{\rm P}_1)$  
& 811.  & 841.  & 0.41  &  0.53  & \\
\\
&                        
& $\chi_{0}(^3{\rm P}_0)$  
& 887.  & 915.  & 0.18  &  0.13 & \\
\\
& $\eta_c(4^1{\rm S}_0) $ 
& $h_c(^1{\rm P}_1)$ 
& 782.  & 808.  & 5.2   &  9.6  & \\
\\
\end{tabular}
\end{ruledtabular}
\label{E1rad_StoP}
\end{table*}

\begin{table*}
\caption{1P and 2P E1 radiative transitions 
(format as in Table~\ref{E1rad_StoP}).}
\vskip 0.3cm
\begin{ruledtabular}
\begin{tabular}{l r r c c c c c c } 
Multiplets 
& Initial meson  & Final meson 
& \multicolumn{2}{c}{E$_{\gamma}$ (MeV)} 
& \multicolumn{2}{c}{$\Gamma_{\rm thy}$~(keV)} 
& $\Gamma_{\rm expt}$~(keV) & \\
&                &          &     NR    & GI   & NR & GI &   \\
\hline
\\
1P $\to$ 1S  
& $\chi_{2}(1^3{\rm P}_2) $ & $J/\psi(1^3{\rm S}_1)$ 
& 429.  & 429.  & 424. & 313.  & 426. $\pm$ 51.   \\
\\
& $\chi_{1}(1^3{\rm P}_1) $ &                        
& 390.  & 389.  & 314. & 239.  & 291. $\pm$ 48.   \\
\\
& $\chi_{0}(1^3{\rm P}_0) $ &                        
& 303.  & 303.  & 152. & 114.  & 119. $\pm$ 19.   \\
\\
& $h_c(1^1{\rm P}_1)$ & $\eta_c(1^1{\rm S}_0)$ 
& 504.  & 496.  & 498. & 352.  & \\ \\
\hline
\\
2P $\to$ 2S  
& $\chi_{2}(2^3{\rm P}_2) $ & $\psi'(2^3{\rm S}_1)$ 
& 276.  & 282.  & 304. & 207.  & \\ \\
& $\chi_{1}(2^3{\rm P}_1) $ &                       
& 232.  & 258.  & 183. & 183.  & \\ \\
& $\chi_{0}(2^3{\rm P}_0) $ &                       
& 162.  & 223.  & 64. & 135.  & \\ \\
& $h_c(2^1{\rm P}_1) $     & $\eta_c'(2^1{\rm S}_0)$ 
& 285.  & 305.  & 280. & 218.  & \\ \\
\hline
\\
2P $\to$ 1S  
& $\chi_{2}(2^3{\rm P}_2) $ & $J/\psi(1^3{\rm S}_1)$ 
& 779.  & 784.  & 81. & 53.  & \\ \\
& $\chi_{1}(2^3{\rm P}_1) $ &                        
& 741.  & 763.  & 71. & 14.  & \\ \\
& $\chi_{0}(2^3{\rm P}_0) $ &                        
& 681.  & 733.  & 56. & 1.3  & \\ \\
& $h_c(2^1{\rm P}_1) $ & $\eta_c(1^1{\rm S}_0)$ 
& 839.  & 856.  & 140. & 85.  & \\ \\
\hline
\\
2P $\to$ 1D  
& $\chi_{2}(2^3{\rm P}_2) $ & $\psi_3(1^3{\rm D}_3)$ 
& 163.  & 128.  & 88. & 29.  & \\ \\
&                           & $\psi_2(1^3{\rm D}_2)$ 
& 168.  & 139.  & 17. & 5.6  & \\ \\
&                           & $\psi(1^3{\rm D}_1)$   
& 197.  & 204.  & 1.9 & 1.0  & \\ \\
& $\chi_{1}(2^3{\rm P}_1) $ & $\psi_2(1^3{\rm D}_2)$ 
& 123.  & 113.  & 35. & 18.  & \\ \\
&                           & $\psi(1^3{\rm D}_1)$   
& 152.  & 179.  & 22. & 21.  & \\ \\
& $\chi_{0}(2^3{\rm P}_0) $ & $\psi(1^3{\rm D}_1)$   
& 81.   & 143.  & 13. & 51.  & \\ \\
& $h_c(2^1{\rm P}_1) $ & $\eta_{2c}(1^1{\rm D}_2)$ 
& 133.  & 117.  & 60. & 27.  & \\ \\
\end{tabular}
\end{ruledtabular}
\label{E1rad_1P2P}
\end{table*}

\begin{table*}
\caption{3P E1 radiative transitions
(format as in Table~\ref{E1rad_StoP}).}
\vskip 0.3cm
\begin{ruledtabular}
\begin{tabular}{l r r c c c c c c }
Multiplets
& Initial meson  & Final meson
& \multicolumn{2}{c}{E$_{\gamma}$ (MeV)}
& \multicolumn{2}{c}{$\Gamma_{\rm thy}$~(keV)}
& $\Gamma_{\rm expt}$~(keV) & \\
&                &          &     NR    & GI   & NR & GI &   \\
\hline
\\
3P $\to$ 3S
& $\chi_{2}(3^3{\rm P}_2) $ & $\psi(3^3{\rm S}_1)$
& 268.  & 231.  & 509. & 199.  & \\ \\
& $\chi_{1}(3^3{\rm P}_1) $ &
& 225.  & 212.  & 303. & 181.  & \\ \\
& $\chi_{0}(3^3{\rm P}_0) $ &
& 159.  & 188.  & 109. & 145.  & \\ \\
& $h_c(3^1{\rm P}_1) $     & $\eta_c(3^1{\rm S}_0)$
& 229.  & 246.  & 276. & 208.  & \\ \\
\hline
\\
3P $\to$ 2S
& $\chi_{2}(3^3{\rm P}_2) $ & $\psi'(2^3{\rm S}_1)$
& 585.  & 602.  & 55. & 30.  & \\ \\
& $\chi_{1}(3^3{\rm P}_1) $ &
& 545.  & 585.  & 45. & 8.9  & \\ \\
& $\chi_{0}(3^3{\rm P}_0) $ &
& 484.  & 563.  & 32. & 0.045  & \\ \\
& $h_c(3^1{\rm P}_1) $     & $\eta_c'(2^1{\rm S}_0)$
& 593.  & 627.  & 75. & 43.  & \\ \\
\hline
\\
3P $\to$ 1S
& $\chi_{2}(3^3{\rm P}_2) $ & $J/\psi(1^3{\rm S}_1)$
& 1048.  & 1063.  & 34. & 19.  &    \\
\\
& $\chi_{1}(3^3{\rm P}_1) $ &
& 1013.  & 1048.  & 31. & 2.2  &    \\
\\
& $\chi_{0}(3^3{\rm P}_0) $ &
& 960.  &  1029.  & 27. & 1.5  &    \\
\\
& $h_c(3^1{\rm P}_1)$ & $\eta_c(1^1{\rm S}_0)$
& 1103.  & 1131.  & 72. & 38.  & \\ \\
\hline
\\
3P $\to$ 2D
& $\chi_{2}(3^3{\rm P}_2) $ & $\psi_3(2^3{\rm D}_3)$
& 147.  & 118.  & 148. & 51.  & \\ \\
&                           & $\psi_2(2^3{\rm D}_2)$
& 156.  & 127.  & 31. &  9.9  & \\ \\
&                           & $\psi(2^3{\rm D}_1)$
& 155.  & 141.  & 2.1 &  0.77  & \\ \\
& $\chi_{1}(3^3{\rm P}_1) $ & $\psi_2(2^3{\rm D}_2)$
& 112.  & 108.  & 58. &  35. & \\ \\
&                           & $\psi(2^3{\rm D}_1)$
& 111.  & 121.  & 19. & 15.  & \\ \\
& $\chi_{0}(3^3{\rm P}_0) $ & $\psi(2^3{\rm D}_1)$
& 43.   &  97.  & 4.4 &  35.  & \\ \\
& $h_c(3^1{\rm P}_1) $ & $\eta_{2c}(2^1{\rm D}_2)$
& 119.  & 109.  & 99. & 48.  & \\ \\
\hline
\\
3P $\to$ 1D
& $\chi_{2}(3^3{\rm P}_2) $ & $\psi_3(1^3{\rm D}_3)$
& 481.  & 461.  & 0.049 & 6.8  & \\ \\
&                           & $\psi_2(1^3{\rm D}_2)$
& 486.  & 470.  & 0.0091 & 0.13  & \\ \\
&                           & $\psi(1^3{\rm D}_1)$
& 512.  & 530.  & 0.00071 & 0.001  & \\ \\
& $\chi_{1}(3^3{\rm P}_1) $ & $\psi_2(1^3{\rm D}_2)$
& 445.  & 452.  & 0.035 & 4.6  & \\ \\
&                           & $\psi(1^3{\rm D}_1)$
& 472.  & 512.  & 0.014 & 0.39  & \\ \\
& $\chi_{0}(3^3{\rm P}_0) $ & $\psi(1^3{\rm D}_1)$
& 410.   & 490.  & 0.037 & 9.7  & \\ \\
& $h_c(3^1{\rm P}_1) $ & $\eta_{2c}(1^1{\rm D}_2)$
& 453.  &  454. & 0.16 & 5.7  & \\ \\
\end{tabular}
\end{ruledtabular}
\label{E1rad_3P}
\end{table*}

\begin{table*}
\caption{1D and 2D E1 radiative transitions
(format as in Table~\ref{E1rad_StoP}).}
\vskip 0.3cm
\begin{ruledtabular}
\begin{tabular}{l r r c c c c c c } 
Multiplets 
& Initial meson  & Final meson 
& \multicolumn{2}{c}{E$_{\gamma}$ (MeV)} 
& \multicolumn{2}{c}{$\Gamma_{\rm thy}$~(keV)} 
& $\Gamma_{\rm expt}$~(keV) & \\
&                &          &     NR    & GI   & NR & GI &   \\
\hline
\\
1D $\to$ 1P  
& $\psi_3(1^3{\rm D}_3) $ & $\chi_{2}(1^3{\rm P}_2)$ 
& 242.  & 282.  & 272. & 296.  & \\ \\
& $\psi_2(1^3{\rm D}_2) $ & $\chi_{2}(1^3{\rm P}_2)$ 
& 236.  & 272.  &  64. & 66.  & \\ \\
&                         & $\chi_{1}(1^3{\rm P}_1)$ 
& 278.  & 314.  & 307. & 268.  & \\ \\
& $\psi(1^3{\rm D}_1) $   & $\chi_{2}(1^3{\rm P}_2)$ 
& 208.  & 208.  &  4.9 & 3.3  & $\leq 330$ $(90\% \ c.l.)$
\cite{Rosner:2004mi,Zhu88}
\\ \\
&                         & $\chi_{1}(1^3{\rm P}_1)$ 
& 250.  & 251.  & 125. & 77.  &  $280 \pm 100 $

\cite{Rosner:2004mi,Zhu88}
\\ \\
&                         & $\chi_{0}(1^3{\rm P}_0)$ 
& 338.  & 338.  & 403. & 213.  & $320 \pm 120$
\cite{Rosner:2004mi,Zhu88}
\\ \\
& $h_{c2}(1^1{\rm D}_2) $ & $h_c(1^1{\rm P}_1)$       
& 264.  & 307.  & 339. & 344.  & \\ \\
\hline
\\
2D $\to$ 2P  
& $\psi_3(2^3{\rm D}_3) $ & $\chi_{2}(2^3{\rm P}_2)$ 
& 190.  &  231.  & 239. & 272.  & \\ \\
& $\psi_2(2^3{\rm D}_2) $ & $\chi_{2}(2^3{\rm P}_2)$ 
& 182.  &  223.  & 52.  & 65. & \\ \\
&                         & $\chi_{1}(2^3{\rm P}_1)$ 
& 226.  &  247.  & 298. & 225.  & \\ \\
& $\psi(2^3{\rm D}_1) $   & $\chi_{2}(2^3{\rm P}_2)$ 
& 183.  & 210.   & 5.9  & 6.3 & \\ \\
&                         & $\chi_{1}(2^3{\rm P}_1)$ 
& 227.  & 234.   & 168. & 114. & \\ \\
&                         & $\chi_{0}(2^3{\rm P}_0)$ 
& 296.  & 269.   & 483. & 191.  & \\ \\
& $\eta_{c2}(2^1{\rm D}_2) $ & $h_c(2^1{\rm P}_1)$       
& 218.  &  244.  & 336. & 296.  & \\ \\
\hline
\\
2D $\to$ 1P  
& $\psi_3(2^3{\rm D}_3) $ & $\chi_{2}(1^3{\rm P}_2)$ 
& 566.  &  609. & 29.   & 16.  & \\ \\
& $\psi_2(2^3{\rm D}_2) $ & $\chi_{2}(1^3{\rm P}_2)$ 
& 558.  &  602. & 7.1  &  0.62 & \\ \\
&                         & $\chi_{1}(1^3{\rm P}_1)$ 
& 597.  &  640. & 26.    & 23.  & \\ \\
& $\psi(2^3{\rm D}_1) $   & $\chi_{2}(1^3{\rm P}_2)$ 
& 559.  & 590. & 0.79    & 0.027 & \\ \\
&                         & $\chi_{1}(1^3{\rm P}_1)$ 
& 598.  &  628. & 14.    & 3.4  & \\ \\
&                         & $\chi_{0}(1^3{\rm P}_0)$ 
& 677.  &  707. & 27.    & 35.  & \\ \\
& $h_{c2}(2^1{\rm D}_2) $ & $h_c(1^1{\rm P}_1)$       
& 585. &  634.  & 40.    & 25.  & \\ \\
\hline
\\
2D $\to$ 1F  
& $\psi_3(2^3{\rm D}_3) $ & $\chi_{4}(1^3{\rm F}_4)$ 
& 143.  &  120.  & 66. & 26.  & \\ \\
&                         & $\chi_{3}(1^3{\rm F}_3)$ 
& 136.  &  114.  & 4.8 & 1.9  & \\ \\
&                         & $\chi_{2}(1^3{\rm F}_2)$ 
& 136.  &  123.  & 14. & 0.055 &\\ \\
& $\psi_2(2^3{\rm D}_2) $ & $\chi_{3}(1^3{\rm F}_3)$ 
& 127.  &  110.  & 44.  & 19. & \\ \\
&                         & $\chi_{2}(1^3{\rm F}_2)$ 
& 127.  &  114.  & 5.6 &  2.4  & \\ \\
& $\psi(2^3{\rm D}_1) $   & $\chi_{2}(1^3{\rm F}_2)$ 
& 128.  &  101.   & 51.  & 17. & \\ \\
& $\eta_{c2}(2^1{\rm D}_2) $ & $h_{c3}(1^1{\rm F}_3)$ 
& 130.  &  112.  &  54. & 23.  & \\ \\
\end{tabular}
\end{ruledtabular}
\label{E1rad_1D2D}
\end{table*}

\begin{table*}
\caption{1F E1 radiative transitions
(format as in Table~\ref{E1rad_StoP}).}
\vskip 0.3cm
\begin{ruledtabular}
\begin{tabular}{l r r c c c c c c } 
Multiplets 
& Initial meson  & Final meson 
& \multicolumn{2}{c}{E$_{\gamma}$ (MeV)} 
& \multicolumn{2}{c}{$\Gamma_{\rm thy}$~(keV)} 
& $\Gamma_{\rm expt}$~(keV) & \\
&                &          &     NR    & GI   & NR & GI &   \\
\hline
\\
1F $\to$ 1D  
& $\chi_4(1^3{\rm F}_4) $ & $\psi_{3}(1^3{\rm D}_3)$ 
& 209.  &  239.  & 332. &  334.  & \\ \\
& $\chi_3(1^3{\rm F}_3) $ & $\psi_{3}(1^3{\rm D}_3)$ 
& 217.  &  240.  &  41. &   38.  & \\ \\
&                         & $\psi_{2}(1^3{\rm D}_2)$ 
& 222.  &  251.  & 354. &  325.  & \\ \\
& $\chi_2(1^3{\rm F}_2) $ & $\psi_{3}(1^3{\rm D}_3)$ 
& 217.  &  236.  &  1.6 &   1.4  & \\ \\
&                         & $\psi_{2}(1^3{\rm D}_2)$ 
& 222.  &  246.  &  62. &   69.  & \\ \\
&                         & $\psi(1^3{\rm D}_1)$     
& 251.  &  309.  & 475. &  541.  & \\ \\
& $h_{c3}(1^1{\rm F}_3) $ & $\eta_{c2}(1^1{\rm D}_2)$& 221.  
&  249.  & 387. &  321.  & \\ \\
\end{tabular}
\end{ruledtabular}
\label{E1rad_1F}
\end{table*}

\begin{table*}
\caption{2F E1 radiative transitions
(format as in Table~\ref{E1rad_StoP}).}
\vskip 0.3cm
\begin{ruledtabular}
\begin{tabular}{l r r c c c c c c } 
Multiplets 
& Initial meson  & Final meson 
& \multicolumn{2}{c}{E$_{\gamma}$ (MeV)} 
& \multicolumn{2}{c}{$\Gamma_{\rm thy}$~(keV)} 
& $\Gamma_{\rm expt}$~(keV) & \\
&                &          &     NR    & GI   & NR & GI &   \\
\hline
\\
2F $\to$ 2D  
& $\chi_4(2^3{\rm F}_4) $ & $\psi_{3}(2^3{\rm D}_3)$ 
& 177.  &  203.  & 307. &  297.  & \\ \\
& $\chi_3(2^3{\rm F}_3) $ & $\psi_{3}(2^3{\rm D}_3)$ 
& 181.  &  204.  &  36. &  35.  & \\ \\
&                         & $\psi_{2}(2^3{\rm D}_2)$ 
& 190.  &  213.  & 334. &  289.  & \\ \\
& $\chi_2(2^3{\rm F}_2) $ & $\psi_{3}(2^3{\rm D}_3)$ 
& 180.  &  200.  &  1.4 &   1.4  & \\ \\
&                         & $\psi_{2}(2^3{\rm D}_2)$ 
& 189.  &  209.  &  58. &   50.  & \\ \\
&                         & $\psi(2^3{\rm D}_1)$     
& 188.  &  222.  & 306. &  295.  & \\ \\
& $h_{c3}(2^1{\rm F}_3) $ & $\eta_{c2}(2^1{\rm D}_2)$
& 188.  &  211.  & 362. &  284.  & \\ \\
\hline
\\
2F $\to$ 1D  
& $\chi_4(2^3{\rm F}_4) $ & $\psi_{3}(1^3{\rm D}_3)$ 
& 508.  &  539.  &  20. &  8.6  & \\ \\
& $\chi_3(2^3{\rm F}_3) $ & $\psi_{3}(1^3{\rm D}_3)$ 
& 512.  &  539.  &  2.3 &  0.14  & \\ \\
&                         & $\psi_{2}(1^3{\rm D}_2)$ 
& 517.  &  549.  &  19. &  11.  & \\ \\
& $\chi_2(2^3{\rm F}_2) $ & $\psi_{3}(1^3{\rm D}_3)$ 
& 511.  &  536.  &  0.090 &  0.003  & \\ \\
&                         & $\psi_{2}(1^3{\rm D}_2)$ 
& 516.  &  545.  &  3.2 &  0.39  & \\ \\
&                         & $\psi(1^3{\rm D}_1)$     
& 542.  &  604.  &  20. &  18.  & \\ \\
& $h_{c3}(2^1{\rm F}_3) $ & $\eta_{c2}(1^1{\rm D}_2)$
& 516.  &  548.  &  22. &  9.9  & \\ \\
\hline
\\
2F $\to$ 1G  
& $\chi_4(2^3{\rm F}_4) $ & $\psi_{5}(1^3{\rm G}_5)$ 
& 132.  &  112.  & 54. &  22.  & \\ \\
&                         & $\psi_{4}(1^3{\rm G}_4)$ 
& 118.  &  104.  & 2.0 &  0.84  & \\ \\
&                         & $\psi_{3}(1^3{\rm G}_3)$ 
& 110.  &  101.  & 0.025 &  0.011  & \\ \\
& $\chi_3(2^3{\rm F}_3) $ & $\psi_{4}(1^3{\rm G}_4)$ 
& 122.  &  105.  &  43. &   18.  & \\ \\
&                         & $\psi_{3}(1^3{\rm G}_3)$ 
& 113.  &  102.  &  2.3 &  1.0  & \\ \\
& $\chi_2(2^3{\rm F}_2) $ & $\psi_{3}(1^3{\rm G}_3)$ 
& 113.  &  98.  &  36. &   16.  & \\ \\
& $h_{c3}(2^1{\rm F}_3) $ & $\eta_{c4}(1^1{\rm G}_4)$
& 123.  &  106.  &  47. &  20.  & \\ \\
\end{tabular}
\end{ruledtabular}
\label{E1rad_2F}
\end{table*}

\begin{table*}
\caption{1G E1 radiative transitions
(format as in Table~\ref{E1rad_StoP}).}
\vskip 0.3cm
\begin{ruledtabular}
\begin{tabular}{l r r c c c c c c } 
Multiplets 
& Initial meson  & Final meson 
& \multicolumn{2}{c}{E$_{\gamma}$ (MeV)} 
& \multicolumn{2}{c}{$\Gamma_{\rm thy}$~(keV)} 
& $\Gamma_{\rm expt}$~(keV) & \\
&                &          &     NR    & GI   & NR & GI &   \\
\hline
\\
1G $\to$ 1F  
& $\psi_5(1^3{\rm G}_5) $ & $\chi_4(1^3{\rm F}_4)$   
& 189.  &  212.  & 373. &  355.  & \\ \\
& $\psi_4(1^3{\rm G}_4) $ & $\chi_4(1^3{\rm F}_4)$   
& 202.  &  219.  &  29. &  25.  & \\ \\
&                         & $\chi_3(1^3{\rm F}_3)$   
& 194.  &  217.  & 382. &  349.  & \\ \\
& $\psi_3(1^3{\rm G}_3) $ & $\chi_4(1^3{\rm F}_4)$   
& 210.  &  222.  &  0.66 &  0.52  & \\ \\
&                         & $\chi_3(1^3{\rm F}_3)$   
& 203.  &  220.  &  37. &   31.  & \\ \\
&                         & $\chi_2(1^3{\rm F}_2)$   
& 203.  &  225.  & 425. &  366.  & \\ \\
& $\eta_{c4}(1^1{\rm G}_4) $ & $h_{c3}(1^1{\rm F}_3)$
& 194.  &  217.  & 407. &  374.  & \\ \\
\end{tabular}
\end{ruledtabular}
\label{E1rad_1G}
\end{table*}

\begin{table*}
\caption{M1 radiative partial widths.
The assumed masses are experimental where known, and otherwise are the theoretical
predictions (see Table~\ref{Table_spectrum}). 
One exception is the $\eta_c(3^1{\rm S}_0)$, for which we assume a mass of 4011~MeV (the 
mass of the known $\psi(4040)$ minus the theoretical 3S splitting). 
We give results for the NR model (both
with and without the recoil factor of $j_0(kr/2)$) and the GI model, which
includes the recoil factor.} 
\vskip 0.3cm
\begin{ruledtabular}
\begin{tabular}{l l l r r c l l l c c }
Initial Multiplet & Initial meson  & Final meson
& \multicolumn{2}{c}{E$_\gamma$ (MeV)\ } & 
& \multicolumn{3}{c}{$\Gamma_{\rm thy}$~(keV)\phantom{sss}  }
& $\Gamma_{\rm expt}$~(keV) & \\
\\
&                &           &   NR    & GI\phantom{x}   & & NR & NR($j_0)$ & GI &   \\
\hline
\\
1S
& $J/\psi(1^3{\rm S}_1) $ & $\eta_c(1^1{\rm S}_0)$   & 116.  &  115. & \phantom{sss} & 2.9  & 2.9  &  2.4  &  $1.1 \pm 0.3$  \\
\\
\hline
\\
2S
& $\psi'(2^3{\rm S}_1) $ & $\eta_c'(2^1{\rm S}_0)$   &  48.  &   48. & & 0.21 & 0.21 &  0.17      & \\
\\
&                        & $\eta_c(1^1{\rm S}_0)$    & 639.  &  638. & & 4.6  & 9.7  &  9.6       & $0.8 \pm 0.2$ \\
\\
& $\eta_c'(2^1{\rm S}_0) $ & $J/\psi(1^3{\rm S}_1)$  & 501.  &  501. & & 7.9  & 2.5  &  5.6       & \\
\\
\hline
\\
3S
& $\psi(3^3{\rm S}_1) $   & $\eta_c(3^1{\rm S}_0)$   &   29. &   35. & & 0.046 &  0.046   & 0.067      & \\
\\
&                         & $\eta_c'(2^1{\rm S}_0)$  &  382. &  436. & & 0.61  &  1.8     & 2.6        & \\
\\
&                         & $\eta_c(1^1{\rm S}_0)$   &  922. &  967. & & 3.5   &  8.7     & 9.0        & \\
\\
& $\eta_c(3^1{\rm S}_0) $ & $\psi'(2^3{\rm S}_1)$    &  312. &  361. & & 1.3   &  0.22    & 0.84       & \\
\\ 
&                         & $J/\psi(1^3{\rm S}_1)$   &  810. &  853. & & 6.3   &  1.9     & 6.9        & \\
\\
\hline
\\
2P
& $h_c'(2^1{\rm P}_1) $   & $\chi_2(1^3{\rm P}_2)$   &  360. &  380.  & &  0.071 &  0.075 & 0.11      & \\
\\
&                         & $\chi_1(1^3{\rm P}_1)$   &  400. &  420.  & &  0.058 &  0.13 & 0.36      & \\
\\
&                         & $\chi_0(1^3{\rm P}_0)$   &  485. &  504.  & &  0.033 &  0.21 & 1.5       & \\
\\
& $\chi_2'(2^3{\rm P}_2)$ & $h_c(1^1{\rm P}_1)$      &  430. &  435.  & &  0.067 &  0.90 & 1.3       & \\
\\
& $\chi_1'(2^3{\rm P}_1)$ & $h_c(1^1{\rm P}_1)$      &  388. &  412.  & &  0.050  &  0.51 & 0.045     & \\
\\
& $\chi_0'(2^3{\rm P}_0)$ & $h_c(1^1{\rm P}_1)$      &  321. &  379.  & &  0.029 &  0.19 & 0.50      & \\
\\
\end{tabular}
\end{ruledtabular}
\label{M1rad}
\end{table*}
\begin{table*}

\caption{Open-flavor strong decays, 3S and 4S states. 
The $\psi(4040)$ and $\psi(4415)$ masses (boldface) are taken from experiment;
for the remaining (unknown) masses we assume the 
theoretical NR potential model values of Table~\ref{Table_spectrum}.}
\vskip 0.3cm
\begin{ruledtabular}
\begin{tabular}{lllll}
Meson 
& State
& Mode 
& $\hskip -0.5cm \Gamma_{thy}$\ \  [$\Gamma_{expt}$] (MeV) 
& Amps. (GeV$^{-1/2}$)
\\
\hline
\hline
$\psi({\bf 4040})$
& $3^3\S_1$ 
& $\D \D$ 
& 0.1 
& $^1$P$_1 = -0.0052$  
\\
&    
& $\D  \D^*$ 
& 33  
& $^3$P$_1 = -0.0954$  
\\
&    
& $\D^* \D^*$ 
& 33   
& $^1$P$_1 = +0.0338$    
\\
&    
& 
&   
& $^5$P$_1 = -0.1510$    
\\
&    
&   
&
& $^5$F$_1 =  0 $    
\\
&    
& $\D_s  \D_s$ 
& 7.8 
& $^1$P$_1 = +0.0518$ 
\\
&    
& {\it total}               
& 74\ $[52 \pm 10]$
&     
\\
\hline
$\eta_c(4043)$
& $3^1\S_0$      
& $\D \D^*$ 
& 47   
& $^3$P$_0 = -0.1139$   
\\
&    
& $\D^* \D^*$ 
& 33 
& $^3$P$_0=-0.1489$   
\\
&   
& {\it total}               
& 80 
&
\\
\hline
\hline
$\psi({\bf 4415})$
& $4^3\S_1$       
& $\D \D$    
& 0.4 
& $^1$P$_1=+0.0066$ 
\\
&    
& $\D \D^*$ 
& 2.3 
&  $^3$P$_1= +0.0177$ 
\\
&    
& $\D^* \D^*$ 
& 16 
& $^1$P$_1= -0.0109$  
\\
&    
&  
&  
& $^5$P$_1= +0.0487$  
\\
&    
&  
&  
& $^5$F$_1= 0$  
\\
&    
& $\D \D_1$ 
& 31  
& $^3$S$_1 = 0$    
\\
&    
& 
&   
& $^3$D$_1 = +0.0933$    
\\
&    
& $\D \D_1'$ 
& 1.0
& $^3$S$_1 = +0.0168$    
\\
&    
& 
&   
& $^3$D$_1 = 0$    
\\
&    
& $\D \D_2^*$  
& 23 
& $^5$D$_1= -0.0881$ 
\\
&    
& $\D^* \D_0^*$ 
& 0.0 
& $^3$S$_1 = -8.7\cdot 10^{-4}$ 
\\
&    
& 
& 0
& $^3$D$_1 = 0$
\\
&    
& $\D_s\D_s$ 
& 1.3 
& $^1$P$_1= -0.0135$  
\\
&    
& $\D_s \D_s^*$ 
& 2.6 
& $^3$P$_1= +0.0212$  
\\
&    
& $\D_s^*  \D_s^*$ 
& 0.7 
& $^1$P$_1 = +0.0027$ 
\\
&    
& 
& 
& $^5$P$_1 = -0.0119$ 
\\
&    
& 
& 
& $^5$F$_1 = 0$ 
\\
&    
& {\it total}               
& 78\ $[43 \pm 15]$
& 
\\
\hline
$\eta_c(4384)$         
& $4^1$S$_0$
&  $\D \D^*$ 
& 6.3 
& $^3$P$_0 = +0.0299$ 
\\
&    
&  $\D^* \D^*$ 
& 14 
& $^3$P$_0 = +0.0473$ 
\\
&    
&  $\D  \D_0^*$ 
& 11 
& $^1$S$_0 = +0.0497$ 
\\
&    
&  $\D  \D_2^*$ 
& 24 
& $^5$D$_0 = -0.1014$ 
\\
&    
&  $\D_s  \D_s^*$ 
& 2.2 
& $^3$P$_0 = +0.0201$ 
\\
&    
&  $\D_s^*  \D_s^*$ 
& 2.2 
& $^3$P$_0 = -0.0231$ 
\\
&    
&  $\D_s \D_{s0}^*$ 
& 0.6 
& $^1$S$_0 = -0.0136$ 
\\
&    
& {\it total}               
& 61 
&  
\\
\end{tabular}
\end{ruledtabular}
\label{strong_3S4S}
\end{table*}

\begin{table*}
\caption{Open-flavor strong decays, 2P and 3P states. The initial 
meson masses are predictions of the NR potential model, 
Table~\ref{Table_spectrum}.}
\vskip 0.3cm
\begin{ruledtabular}
\begin{tabular}{lllll}
Meson 
& State
& Mode 
& \hskip -0.5cm $\Gamma_{thy}$ (MeV) 
& Amps. (GeV$^{-1/2}$)
\\
\hline
\hline
$\chi_2(3972)$         
&
$2^3$P$_2$         
& $\D \D$ 
& 42 
& $^1$D$_2 = +0.0992$ 
\\
&                           
& $\D \D^*$ 
& 37 
& $^3$D$_2 = -0.1172$ 
\\
&                          
& $\D_s \D_s$ 
& 0.7 
& $^1$D$_2 = +0.0202$ 
\\
&                   
& {\it total}               
&  80 
&
\\
\hline
$\chi_1(3925)$ 
& $2^3$P$_1$
& $\D \D^*$ 
& 165 
& $^3$S$_1 = +0.2883$    
\\
&
& 
&   
& $^3$D$_1 = -0.0525$    
\\
\hline
$\chi_0(3852)$ 
& $2^3$P$_0$
& $\D \D$ 
& 30 
& $^1$S$_0 = +0.1025$ 
\\
\hline
$h_c(3934)$
& $2^1$P$_1$ 
& $\D \D^*$ 
& 87 
& $^3$S$_1 = -0.1847$    
\\
&
& 
&   
& $^3$D$_1 = -0.0851$    
\\
\hline
\hline
$\chi_2(4317)$  
& $3^3\P_2$
& $\D \D $ 
&  8.0  
& $^1\D_2 = -0.0330$  
\\
&      
& $\D   \D^*$ 
& 2.4 
& $^3\D_2 = +0.0191$ 
\\
&
& $\D^* \D^*$ 
& 24 
& $^5\S_2 = +0.0592$  
\\
&                 
&  
&  
& $^1\D_2 = +0.0107$ 
\\
&                 
&  
&  
& $^5\D_2 = -0.0282$ 
\\
&                 
&                 
&  
& $^5\G_2 = 0$ 
\\
&
& $\D  \D_1$  
& 1.1 
& $^3\P_2 = +0.0240$ 
\\
&                 
& 
&  
& $^3\F_2 = +0.0105$ 
\\
&                 
& $\D  \D_1'$  
& 12 
& $^3\P_2 = -0.0915$ 
\\
&                 
& 
&  
& $^3\F_2 = 0$ 
\\
&                 
& $\D_s \D_s$ 
& 0.8 
& $^1\D_2 = +0.0115$ 
\\
&                 
& $\D_s    \D_s^*$ 
& 11 
& $^3\D_2 = -0.0474$ 
\\
&                 
& $\D_s^*  \D_s^*$ 
& 7.2 
& $^5\S_2 = -0.0266$  
\\
&                 
&                 
& 
& $^1\D_2 = +0.0145$ 
\\
&                 
& 
& 
& $^5\D_2 = -0.0384$ 
\\
&                 
&  
&  
& $^5\G_2 = 0$ 
\\
&
& {\it total} 
& 66  
& 
\\
\hline
$\chi_1(4271)$  
& $3^3\P_1$
& $\D     \D^*$ 
& 6.8 
& $^3\S_1 = -0.0337$ 
\\
&      
& 
&  
& $^3\D_1 = +0.0011$ 
\\
&
& $\D^*   \D^*$ 
& 19 
& $^5\D_1 = -0.0632$ 
\\
&
& $\D     \D_0^*$ 
& 0.1 
& $^1\P_1 = -0.0062$ 
\\
&                 
& $\D_s   \D_s^*$ 
& 9.7 
& $^3\S_1 = -0.0287$ 
\\
&
& 
&  
& $^3\D_1 = -0.0385$ 
\\
&
& $\D_s^* \D_s^*$ 
& 2.7 
& $^5\D_1 = -0.0356$ 
\\
&
& {\it total} 
& 39  
& 
\\
\hline
$\chi_0(4202)$ 
& $3^3\P_0$
& $\D \D$ 
& 0.5 
& $^1\S_0 = -0.0091$ 
\\
&
& $\D^* \D^*$ 
& 43 
& $^1\S_0 = -0.0267$ 
\\
&
& 
&  
& $^5\D_0 = -0.0997$ 
\\
&
& $\D_s \D_s$ 
& 6.8
& $^1\S_0 = -0.0374$ 
\\
&
& {\it total} 
& 51  
& 
\\
\hline
$h_{c}(4279)$  
& $3^1\P_1$
& $\D \D^*$ 
& 3.0 
& $^3\S_1 = +0.0216$
\\
&
&
&
& $^3\D_1 = +0.0048$ 
\\
&
& $\D^*  \D^*$ 
& 22 
& $^3\S_1 = +0.0456$ 
\\
&
&
&
& $^3\D_1 = -0.0487$ 
\\
&                
& $\D  \D_0^*$ 
& 28 
& $^1\P_1 = -0.0943$ 
\\
&                
& $\D_s \D_s^*$ 
& 15 
& $^3\S_1 = +0.0222$  
\\
&
&
&
& $^3\D_1 = -0.0539$ 
\\
&                
& $\D_s^*  \D_s^*$  
& 7.5 
& $^3\S_1 = -0.0464$  
\\
&
& 
&  
& $^3\D_1 = -0.0327$ 
\\
&                
& {\it total}
& 75 
& 
\\
\end{tabular}
\end{ruledtabular}
\label{strong_2P3P}
\end{table*}

\begin{table*}
\caption{Open-flavor strong decays, 1D and 2D states. 
The $\psi(3770)$ and $\psi(4159)$ masses (boldface) are taken from experiment;
for the remaining (unknown) masses we assume the 
theoretical NR potential model values of Table~\ref{Table_spectrum}.}
\vskip 0.3cm
\begin{ruledtabular}
\begin{tabular}{lllll}
Meson 
& State
& Mode 
& $\hskip -0.5cm \Gamma_{thy}$\ \  [$\Gamma_{expt}$] (MeV) 
& Amps. (GeV$^{-1/2}$)
\\
\hline
\hline
$\psi_3(3806)$ 
& $1^3\D_3$ 
& DD 
& 0.5
& $^1$F$_3 = +0.0150$  
\\
\hline
$\psi({\bf 3770})$
& $1^3\D_1$ 
& DD 
& 43\  [$ 23.6 \pm 2.7$]
& $^1$P$_1 = +0.1668$  
\\
\hline
\hline
$\psi_3(4167)$     
& $2^3\D_3$ 
& $\D   \D$ 
& 24  
& $^1\F_3 = +0.0631$  
\\             
&
& $\D \D^*$ 
& 50 
& $^3\F_3 = -0.0997$ 
\\
&
& $\D^* \D^*$ 
& 67 
& $^5\P_3 = -0.1249$
\\
&
&  
&  
& $^1\F_3 = +0.0218$  
\\
&
&  
&  
& $^5\F_3 = -0.0478$ 
\\
&
&  
&  
& $^5\H_3 = 0$ 
\\
&
& $\D_s \D_s$ 
& 5.7 
& $^1\F_3 = +0.0358$ 
\\
&
& $\D_s \D_s^*$
& 1.2 
& $^3\F_3 = -0.0205$ 
\\
&
& {\it total}               
& 148
&     
\\
\hline
$\psi_2(4158)$     
& $2^3\D_2$ 
& $\D  \D^*$  
& 34 
& $^3\P_2= +0.0121$  
\\
&
&
&                     
& $^3\F_2=-0.0822$ 
\\
&
& $\D^* \D^*$ 
& 32 
& $^5\P_2 =  -0.0660$  
\\
&
&
&
& $^5\F_2 =  -0.0685$ 
\\
&
& $\D_s \D_s^*$ 
& 26 
& $^3\P_2 =  +0.0983$  
\\
&
&
&                     
& $^3\F_2 = -0.0149$ 
\\
&
& {\it total}               
& 92
&     
\\
\hline
$\psi({\bf 4159})$       
& $2^3{\D}_1$       
& $\D \D$ 
& 16 
& $^1\P_1 =  -0.0522$ 
\\
&
& $\D \D^*$  
& 0.4 
& $^3\P_1 =  +0.0085$ 
\\
&
& $\D^* \D^*$  
& 35 
& $^1\P_1 = +0.0489$
\\
&
&
&                     
& $^5\P_1 = -0.0219$ 
\\
&
&
&               
& $^5\F_1 = -0.0845$ 
\\
&
& $\D_s \D_s$  
& 8.0 
& $^1\P_1 = -0.0427$ 
\\
&
& $\D_s \D_s^*$  
& 14 
& $^3\P_1 = +0.0733$ 
\\
&
& {\it total}
& 74\  [$ 78 \pm 20$]
&
\\
\hline
$\eta_{c2}(4158)$  
& $2^1{\D}_2$  
& $\D \D^*$ 
& 50 
& $^3\P_2 = -0.0099$  
\\
&
&
&                     
& $^3\F_2= -0.1007$ 
\\
&
& $\D^* \D^*$ 
& 43 
& $^3\P_2 = -0.0933$  
\\
&
&
&                     
& $^3\F_2 = -0.0593$ 
\\
&
& $\D_s \D_s^*$ 
& 18 
& $^3\P_2 = -0.0802$ 
\\
&
&
&
& $^3\F_2 = -0.0182$ 
\\
&
& {\it total}
& 111
&
\\
\end{tabular}
\end{ruledtabular}
\label{strong_1D2D}
\end{table*}

\begin{table*}
\caption{Open-flavor strong decays, 1F states. The initial meson
masses are predictions of the NR potential model, Table~\ref{Table_spectrum}.}
\vskip 0.3cm
\begin{ruledtabular}
\begin{tabular}{lllll}
Meson
& State
& Mode
& $\hskip -0.5cm \Gamma_{thy}$ (MeV)
& Amps. (GeV$^{-1/2}$)
\\
\hline
\hline
$\chi_4(4021)$
& $1^3{\F}_4$
& $\D \D$
& 6.8
& $^1\G_4 = + 0.0379$
\\
&
& $\D \D^*$
& 1.4
& $^3\G_4 = - 0.0204$
\\
&
& $\D^* \D^*$
& 0.05
& $^5\D_4 = - 0.0092$
\\
&
&
&
& $^1\G_4 = + 1.2\cdot 10^{-5}$
\\
&
&
&
& $^5\G_4 = - 2.3\cdot 10^{-5}$
\\
&
&
&
& $^5\I_4 = 0$
\\
&
& $\D_s\D_s$
& 0.02
& $^1\G_4 = + 0.0029$
\\
&
& {\it total}
& 8.3
&
\\
\hline
$\chi_3(4029)$
& $1^3{\F}_3$
& $\D \D^*$
& 83
& $^3\D_3= + 0.1534$
\\
&
&
&
& $^3\G_3= - 0.0197$
\\
&
& $\D^*\D^*$
& 0.2
& $^5\D_3 = - 0.0136$
\\
&
&
&
& $^5\G_3 = - 2.4\cdot 10^{-4}$
\\
&
& {\it total}
& 84
&
\\
\hline
$\chi_2(4029)$
& $1^3{\F}_2$
& $\D \D$
& 98
& $^1\D_2 = + 0.1430$
\\
&
& $\D \D^*$
& 57
& $^3\D_2 = + 0.1283$
\\
&
& $\D^* \D^*$
& 0.1
& $^5\S_2 = 0$
\\
&
&
&
& $^1\D_2 = + 0.0080$
\\
&
&
&
& $^5\D_2= - 0.0061$
\\
&
&
&
& $^5\G_2= - 3.1\cdot 10^{-4}$
\\
&
& $\D_s \D_s$
& 5.9
& $^1\D_2 = + 0.0464$
\\
&
& {\it total}
& 161
&
\\
\hline
$h_{c3}(4026)$
& $1^1{\F}_3$
& $\D \D^*$
& 61
& $^3\D_3 = - 0.1313$
\\
&
&
&
& $^3\G_3= - 0.0220$
\\
&
& $\D^* \D^*$
& 0.1
& $^3\D_3 = - 0.0129$
\\
&
&
&
& $^3\G_3= - 1.3 \cdot 10^{-4}$
\\
&
& {\it total}
& 61
&
\\
\end{tabular}
\end{ruledtabular}
\label{strong_1F}
\end{table*}

\begin{table*}
\caption{Open-flavor strong decays, 2F states. The initial meson masses
are predictions of the NR potential model, Table~\ref{Table_spectrum}.}
\vskip 0.3cm
\begin{ruledtabular}
\begin{tabular}{lllll}
Meson 
& State
& Mode 
& $\hskip -0.5cm \Gamma_{thy}$ (MeV) 
& Amps. (GeV$^{-1/2}$)
\\
\hline
\hline
$\chi_4(4348)$     
& $2^3\F_4$ 
& $\D \D$ 
& 12 
& $^1\G_4 = + 0.0395$  
\\
&
& $\D \D^*$ 
& 31 
& $^3\G_4 = - 0.0679$ 
\\
&
& $\D^* \D^*$ 
& 21 
& $^5\D_4 = - 0.0386$ 
\\
&
&
&                     
& $^1\G_4 = + 0.0209$ 
\\
&
&
&                     
& $^5\G_4 = - 0.0415$
\\
&
&
&                     
& $^5\I_4 = 0 $
\\
&
& $ \D \D_1$ 
& 0.5  
& $^3\F_4 = + 0.0082$  
\\
&
&
&                     
& $^3\H_4 = + 0.0015$ 
\\
&
& $ \D \D_1'$ 
& 2.0
& $^3\F_4 = - 0.0172$ 
\\
&
&
&                     
& $^3\H_4 = 0$ 
\\
&
& $\D \D_2^*$ 
& 0.04 
& $^5\F_4 = - 0.0051$ 
\\
&
&
&                     
& $^3\H_4 = - 1.1 \cdot 10^{-4}$ 
\\
&
& $\D^* \D_0^*$ 
& 0.06 
& $^3\F_4 = - 0.0058$ 
\\
&
&
&                     
& $^3\H_4 = 0$ 
\\
&
& $\D_s \D_s$ 
& 5.0 
& $^1\G_4 = + 0.0283$ 
\\
&
& $\D_s \D_s^*$ 
& 4.3 
& $^3\G_4 = - 0.0290$ 
\\
&
& $\D_s^* \D_s^*$ 
& 11 
& $^5\D_4 = - 0.0562$ 
\\
&
&
&                     
& $^1\G_4 = + 0.0034$  
\\
&
&
&                     
& $^5\G_4 = - 0.0068$ 
\\
&
&
&                     
& $^5\I_4 = 0$
\\
&
& {\it total} 
&  87
& 
\\
\hline
$\chi_3(4352)$     
& $2^3\F_3$     
& $\D\D^*$  
& 27 
& $^3\D_3 = - 0.0208$  
\\
&
&
&
& $^3\G_3 = - 0.0597$ 
\\
&
& $\D^* \D^*$ 
& 27 
& $^5\D_3 = - 0.0212$  
\\
&
&
&
& $^5\G_3 = - 0.0656$ 
\\
&
& $\D \D_0^*$ 
& 0.6 
& $^1\F_3 = -0.0119$ 
\\
&
& $\D\D_1$ 
& 3.4 
& $^3\F_3 = - 0.0237$ 
\\
&
& $\D\D_1'$ 
& 1.3 
& $^3\F_3 = - 0.0147$ 
\\
&
& $\D \D_2^*$ 
& 32 
& $^5\P_3 = + 0.1432$  
\\
&
&
&
& $^5\F_3= + 0.0025$ 
\\
&
&
&
& $^5\H_3= - 1.9 \cdot 10^{-4} $ 
\\
&
& $\D^* \D_0^*$ 
&  0.2
& $^3\F_3 = - 0.0091$  
\\
&
& $\D_s \D_s^*$ 
& 13 
& $^3\D_3 = + 0.0434$  
\\
&
&
&
& $^3\G_3 = - 0.0260$ 
\\
&
& $\D_s^* \D_s^*$ 
& 4.3 
& $^5\D_3 = - 0.0328$ 
\\
&
&
&
& $^5\G_3 = - 0.0112$ 
\\
&
& $\D_s\D_{s0}^*$ 
& 0.04 
& $^1\F_3 = - 0.0039$ 
\\
&
& {\it total} 
& 110 
& 
\\
\hline
$\chi_2(4351)$     
& $2^3\F_2$     
& $\D \D$ 
& 15 
& $^1\D_2 = - 0.0446$ 
\\
&                   
& $\D \D^*$ 
& 2.0
& $^3\D_2 = - 0.0171$ 
\\
&                    
& $\D^* \D^*$ 
& 41 
& $^5\S_2 = 0$  
\\
&                   
&                   
&                   
& $^1\D_2 = + 0.0127$  
\\
&                   
&                   
&                   
& $^5\D_2 = - 0.0096$  
\\
&                   
&                   
&                   
& $^5\G_2 = - 0.0828$ 
\\
&                    
& $\D \D_1$ 
& 105  
& $^3\P_2 = + 0.2093$ 
\\
&
&
& 
& $^3\F_2 = + 0.0167$ 
\\
&               
& $\D \D_1'$ 
& 0.3  
& $^3\P_2 = 0 $ 
\\
&
&
& 
& $^3\F_2 = - 0.0113$ 
\\
&                    
& $\D \D_2^*$ 
& 5.5 
& $^5\P_2 = + 0.0592$ 
\\
&                    
& $\D^* \D_0^*$ 
& 0.2 
& $^3\F_2 = + 0.0064 $ 
\\
&                   
& $\D_s \D_s$ 
& 1.1 
& $^1\D_2 = + 0.0134$ 
\\
&                     
& $\D_s \D_s^*$ 
&6.9
& $^3\D_2 = + 0.0365$ 
\\
&                   
& $\D_s^* \D_s^*$ 
& 2.7 
& $^5\S_2 = 0$  
\\
&
&
&
& $^1\D_2 = + 0.0194$  
\\
&
&
&
& $^5\D_2 = - 0.0146$  
\\
&
&
&
& $^5\G_2 = - 0.0140$ 
\\
&
& {\it total} 
& 180 
& 
\\
\end{tabular}
\end{ruledtabular}
\label{strong_2Fa}
\end{table*}

\setcounter{table}{13}

\begin{table*}
\caption{Open-flavor strong decays, 2F states (cont.).}
\vskip 0.3cm
\begin{ruledtabular}
\begin{tabular}{lllll}
Meson 
& State
& Mode 
& $\hskip -0.5cm \Gamma_{thy}$ (MeV) 
& Amps. (GeV$^{-1/2}$)
\\
\hline
\hline
$h_{c3}(4350)$     
& $2^1\F_3$ 
& $\D \D^*$ 
& 34 
& $^3\D_3 = + 0.0173$   
\\
&
&
&
& $^3\G_3 = - 0.0689$  
\\
&
& $\D^* \D^*$ 
& 24 
& $^3\D_3 = - 0.0267$ 
\\
&
&
&
& $^3\G_3 = - 0.0584$ 
\\
&
& $\D \D_0^*$ 
& 11 
& $^1\F_3 = - 0.0516$ 
\\
&
& $\D \D_1$ 
& 0.02  
& $^3\F_3 = - 0.0026$ 
\\
&
& $\D \D_1'$ 
& 0.03  
& $^3\F_3 = - 0.0035$ 
\\
&
& $\D \D_2^*$ 
& 22 
& $^5\P_3 = - 0.1209$
\\
&
&
&
& $^5\F_3 = - 0.0070$ 
\\
&
&
&
& $^5\H_3 = - 1.8 \cdot 10^{-4} $
\\
&
& $\D^* \D_0^*$ 
& 0.006
& $^3\F_3 = - 0.0018 $
\\
&
& $\D_s \D_s^*$ 
& 12 
& $^3\D_3 = -0.0379$  
\\
&
&
&
& $^3\G_3 = -0.0297$ 
\\
&
& $\D_s^* \D_s^*$ 
& 6.0 
& $^3\D_3 = - 0.0400$  
\\
&
&
&
& $^3\G_3 = - 0.0098$ 
\\
&
& $\D_s\D_{s0}^*$ 
& 0.3  
& $^1\F_3 = - 0.0108$ 
\\
&
& {\it total} 
&  109
& 
\\
\end{tabular}
\end{ruledtabular}
\label{strong_2Fb}
\end{table*}

\begin{table*}
\caption{Open-flavor strong decays, 1G states. The initial meson masses
are predictions of the NR potential model, Table~\ref{Table_spectrum}.}
\vskip 0.3cm
\begin{ruledtabular}
\begin{tabular}{lllll}
Meson 
& State
& Mode 
& $\hskip -0.5cm \Gamma_{thy}$ (MeV) 
& Amps. (GeV$^{-1/2}$)
\\
\hline
\hline
$\psi_5(4214)$     
& $1^3\G_5$
& $\D \D$  
& 10 
& $^1\H_5 = + 0.0397$ 
\\
&         
& $\D \D^*$  
& 6.4 
& $^3\H_5 = - 0.0342$ 
\\
&
& $\D^* \D^*$  
& 41 
& $^5\F_5 = - 0.0982$ 
\\
&
&
&                     
& $^1\H_5 = + 0.0045$ 
\\
&
&
&                    
& $^5\H_5 = - 0.0083$ 
\\
&
&
&                    
& $^5\J_5 = 0$ 
\\
&                    
& $\D_s \D_s$  
& 0.4 
& $^1\H_5 = + 0.0086$ 
\\
&
& $\D_s \D_s^*$  
& 0.0 
& $^3\H_5 = - 0.0027$ 
\\
&                    
& {\it total} 
& 58
&
\\
\hline
$\psi_4(4228)$     
& $1^3\G_4$
& $\D \D^*$  
& 88 
& $^3\F_4 = + 0.1209$  
\\
&
&
&                   
& $^3\H_4 = - 0.0334$ 
\\
&
& $\D^* \D^*$ 
& 19 
& $^5\F_4 = - 0.0643$  
\\
&
&
&                   
& $^5\H_4 = - 0.0153$ 
\\
&
& $\D_s \D_s^*$ 
& 3.5 
& $^3\F_4 = + 0.0307$  
\\
&
&
&                   
& $^3\H_4 = - 0.0031$ 
\\
&
& $\D_s^* \D_s^*$ 
& 0.0 
& $^5\F_4 = - 1.5 \cdot 10^{-4}$  
\\
&
&
&                   
& $^5\H_4 = - 6.0\cdot 10^{-7}$ 
\\
&
& {\it total} 
& 111
&
\\
\hline
$\psi_3(4237)$     
& $1^3\G_3$
& $\D\D$ 
& 63 
& $^1\F_3 = + 0.0974$ 
\\
&
& $\D \D^*$ 
& 66 
& $^3\F_3 = + 0.1077$ 
\\
&
& $\D^* \D^*$ 
& 13 
& $^1\F_3 = + 0.0358$
\\
&
&
&                     
& $^5\F_3 = - 0.0326$ 
\\
&
&
&
& $^5\H_3=  - 0.0217$ 
\\
&
& $\D_s \D_s$ 
& 10 
& $^1\F_3 = + 0.0443$ 
\\
&
& $\D_s \D_s^*$ 
& 3 
& $^3\F_3 = + 0.0289$ 
\\
&
& $\D_s^* \D_s^*$ 
& 0.0 
& $^1\F_3 = + 4.6 \cdot 10^{-4}$  
\\
&
&
&
& $^5\F_3 = - 4.2 \cdot 10^{-4}$  
\\
&
&
&
& $^5\H_3= - 1.4\cdot 10^{-5}$ 
\\
&
& {\it total} 
& 155
&
\\
\hline
$\eta_{c4}(4225)$  
& $1^1\G_4$
& $\D \D^*$  
& 72 
& $^3\F_4 = - 0.1076$ 
\\
&
&
&                     
& $^3\H_4 = - 0.0367$
\\
&
& $\D^* \D^*$ 
& 24 
& $^3\F_4 = - 0.0732$  
\\
&
&
&                     
& $^3\H_4 = - 0.0136$ 
\\
&
& $\D_s \D_s^*$ 
& 3 
& $^3\F_4 = - 0.0268$  
\\
&
&
&                     
& $^3\H_4 = - 0.0033$ 
\\
&
& $\D_s^* \D_s^*$ 
& 0.0 
& $^3\F_4 = -2.2\cdot10^{-5}$  
\\
&
&
&
& $^3\H_4 = -1.7\cdot 10^{-8}$ 
\\
&
& {\it total} 
&99 
&
\\
\end{tabular}
\end{ruledtabular}
\label{strong_1G}
\end{table*}

\end{document}